\documentclass[11pt,preprint]{aastex}

\usepackage{natbib}
\usepackage{ulem}
\usepackage{lineno}

\usepackage{lscape}
\usepackage{rotating}
\usepackage{float}
\usepackage{graphicx}
\usepackage{epstopdf}
\usepackage{longtable}

\usepackage{verbatim}

\newcommand{\pflux}{ph cm$^{-2}$ s$^{-1}$}
\newcommand{\Fermi}{\textit{Fermi}}

\linenumbers

\shorttitle{{\it Fermi}-LAT Unassociated Sources}
\shortauthors{Abdo et al.}

\begin{document}

\title{A Statistical Approach to Recognizing Source Classes for Unassociated Sources in
the First Fermi-LAT Catalog}


\author{
M.~Ackermann\altaffilmark{2}, 
M.~Ajello\altaffilmark{3}, 
A.~Allafort\altaffilmark{3}, 
E.~Antolini\altaffilmark{4,5}, 
L.~Baldini\altaffilmark{6}, 
J.~Ballet\altaffilmark{7}, 
G.~Barbiellini\altaffilmark{8,9}, 
D.~Bastieri\altaffilmark{10,11}, 
R.~Bellazzini\altaffilmark{6}, 
B.~Berenji\altaffilmark{3}
% ...
}
% ...
\altaffiltext{2}{Deutsches Elektronen Synchrotron DESY, D-15738 Zeuthen, Germany}
\altaffiltext{3}{W. W. Hansen Experimental Physics Laboratory, Kavli Institute for Particle Astrophysics and Cosmology, Department of Physics and SLAC National Accelerator Laboratory, Stanford University, Stanford, CA 94305, USA}
\altaffiltext{4}{Istituto Nazionale di Fisica Nucleare, Sezione di Perugia, I-06123 Perugia, Italy}
\altaffiltext{5}{Dipartimento di Fisica, Universit\`a degli Studi di Perugia, I-06123 Perugia, Italy}
\altaffiltext{6}{Istituto Nazionale di Fisica Nucleare, Sezione di Pisa, I-56127 Pisa, Italy}
\altaffiltext{7}{Laboratoire AIM, CEA-IRFU/CNRS/Universit\'e Paris Diderot, Service d'Astrophysique, CEA Saclay, 91191 Gif sur Yvette, France}
\altaffiltext{8}{Istituto Nazionale di Fisica Nucleare, Sezione di Trieste, I-34127 Trieste, Italy}
\altaffiltext{9}{Dipartimento di Fisica, Universit\`a di Trieste, I-34127 Trieste, Italy}
\altaffiltext{10}{Istituto Nazionale di Fisica Nucleare, Sezione di Padova, I-35131 Padova, Italy}
\altaffiltext{11}{Dipartimento di Fisica ``G. Galilei", Universit\`a di Padova, I-35131 Padova, Italy}
% ...

\begin{abstract}
The \Fermi~Large Area Telescope First Source Catalog (1FGL) provided spatial, spectral, and temporal properties for a large number of $\gamma$-ray sources using a uniform analysis method. After correlating with the most-complete catalogs of source types known to emit $\gamma$ rays, 630 of these sources are ``unassociated'' (i.e. have no obvious counterparts at other wavelengths). Here, we employ two statistical analyses of the primary $\gamma$-ray characteristics for these unassociated sources in an effort to correlate their $\gamma$-ray properties with the AGN and pulsar populations in 1FGL. Based on the correlation results, we classify 221 AGN-like and 134 pulsar-like sources in the 1FGL unassociated sources. The results of these source ``classifications'' appear to match the expected source distributions, especially at high Galactic latitudes. While useful for planning future multiwavelength follow-up observations, these analyses use limited inputs, and their predictions should not be considered equivalent to ``probable source classes'' for these sources. We discuss multiwavelength results and catalog cross-correlations to date, and provide new source associations for 229 \Fermi-LAT sources that had no association listed in the 1FGL catalog. By validating the source classifications against these new associations, we find that the new association matches the predicted source class in $\sim$80\% of the sources. 


\end{abstract}




\section{\label{sec:intro}Introduction}

Astrophysical sources of high-energy $\gamma$ rays (photon energies above 10 MeV), although inherently interesting as tracers of energetic processes in the Universe, have long been hard to identify.  Only four of the 25 sources in the second COS-B catalog had identifications \citep{COSBcatalog}, and over half the sources in the third EGRET catalog had no associations with known objects \citep{3rdCat}.  A principal reason for the difficulty of finding counterparts to high-energy $\gamma$-ray sources has been the large positional errors in their measured locations, a result of the limited photon statistics and angular resolution of the $\gamma$-ray observations and the bright diffuse $\gamma$-ray emission from the Milky Way. In addition, a number of the COS-B and EGRET sources were determined to be spurious by follow-up analysis and observations.

A major step forward for detection and identification of high-energy $\gamma$-ray sources came when the {\it Gamma-ray Large Area Space Telescope} (GLAST) was launched on 2008 June 11. It began its scientific operations two months later, and shortly thereafter, it was renamed the \Fermi~Gamma-ray Space Telescope.  Its primary instrument is the Large Area Telescope \citep[LAT;][]{LATinstrument}, the successor to the Energetic Gamma-Ray Experiment Telescope (EGRET) on  the {\it Compton Gamma-Ray Observatory} \citep{Thompson93}.  The LAT offers a major increase in sensitivity over EGRET, allowing it to study the 100 MeV to $\sim$300 GeV $\gamma$-ray sky in unprecedented detail. 

The high sensitivity, improved angular resolution and nearly uniform sky coverage of the LAT make it a powerful tool for detecting and characterizing large numbers of $\gamma$-ray sources.  The \Fermi-LAT First Source Catalog \citep[1FGL;][]{1FGL} lists 1451 sources detected during the first 11 months of operation by the LAT, of which 821 were shown to be associated with at least one plausible counterpart. Of these, 698 were extragalactic (mostly Active Galactic Nuclei, or AGNs) and 123 were Galactic (mostly pulsars and supernova remnants, but also pulsar wind nebulae and high-mass X-ray binaries). After the publication of the 1FGL catalog, the association panorama evolved very quickly with the release of the catalog of AGNs \citep[1LAC;][]{1LAC} as well as a catalog of pulsar wind nebulae (PWNe) and supernova remnants (SNRs) \citep{LAT_PwnSnrSearch11}. 


Here, as a starting point for our multivariate classification strategy, we consider the entire original list of 630 1FGL sources that remain unassociated with plausible counterparts at other wavelengths. A plausible counterpart is a member of a known or likely $\gamma$-ray emitting class located close to the 95\% uncertainty radius of a given 1FGL source, with an association confidence of 80\% or higher \citep{1FGL}. The 95\% uncertainty radii for 1FGL source locations are typically 10$^{\prime}$.  While greatly improved over the degree-scale uncertainties of previous instruments, these position measurements are still inadequate to make firm identifications based solely on location.

We have taken a multi-pronged approach toward understanding these unassociated 1FGL sources, using all the available information about the $\gamma$-ray sources.  Information about locations, spectra, and variability has been combined with properties of the established $\gamma$-ray source classes and multiwavelength counterpart searches.

Here we look in depth at the properties of the 1FGL unassociated sources, and investigate the implications of those characteristics. Specifically, this paper addresses five primary questions:

\begin{enumerate}
\item What do the $\gamma$-ray properties of the unassociated 1FGL sources reveal about these sources (Section~\ref{sec:sample})? 
\item What does our understanding of the $\gamma$-ray properties of the associated sources suggest about the possible source class for each of the 1FGL unassociated sources  (Section~\ref{sec:classification})?
\item What new associations or multiwavelength counterparts have been found beyond those from the first LAT catalog (Section~\ref{sec:newassoc})?
\item Do the new classifications properly predict sources that have been associated since the release of the 1FGL catalog (Section~\ref{sec:discuss})?
\item What do the new classifications and associations imply about the existence of unknown new $\gamma$-ray source classes (Section~\ref{sec:conclusion})?

\end{enumerate}

Although the 2FGL catalog \citep{2FGL} was being developed in parallel with the present work, we focus on the 1FGL results, where some follow-up results are available for comparison with the methods of this work.  Such follow-up observations for 2FGL have yet  to be done.

\section{\label{sec:sample}Gamma-ray properties of unassociated \Fermi-LAT sources}

In the 1FGL catalog \citep[][hereafter ``1FGL'']{1FGL}, source identifications and associations were assigned through an objective procedure. For a source to be considered identified in the 1FGL catalog, detection of periodic emission (pulsars or X-ray binaries) or variability correlated with observations at other wavelengths (blazars) was required. Additionally, measurement of an angular extent consistent with observations at other wavelengths was used to declare identifications for a few sources associated with SNRs and radio galaxies \citep{LATW51C,LATCenA,LAT_W44,LATIC443}.  Associations were reported only for sources with  positional correlations between LAT sources and members of plausible source classes (based on Bayesian probabilities of finding a source of a given type in a LAT error box). This automated procedure was based on a list of 32 catalogs that contain potential counterparts of LAT sources based either on prior knowledge about classes of high-energy $\gamma$-ray emitters or on theoretical expectations. In addition, it indicated coincident detections at radio frequencies and TeV energies, as well as positional coincidences with EGRET and {\it AGILE} sources.

In total 821 of the 1451 sources in the 1FGL catalog (56\%) were associated with a least one counterpart by the automated procedure, with 779 being associated using the Bayesian method while 42 are spatially correlated with extended sources based on overlap of the error regions and source extents. From the simulations in 1FGL we expect that $\sim$57 among the 821 sources (7\%) are associated spuriously in 1FGL.  We found the initial list of unassociated sources by simply extracting the list of 1FGL sources without any association from the 1FGL catalog. These sources are spread across the sky, with about 40\% located within $10\degr$ of the Galactic plane. 

Sources without firm identifications that are in regions of enhanced diffuse $\gamma$-ray emission along the Galactic plane or are near local interstellar cloud complexes (like Orion), sources that lie along the Galactic ridge ($300\degr<l<60\degr$, $|b|<1\degr$), and sources that are in regions with source densities great enough that their  position error estimates overlap in the $\gamma$-ray data are called c-sources, as their 1FGL designator has a ``c'' appended to indicate ``caution'' or ``confused region''. The remainder of the unassociated sources did not have a ``caution'' designator in 1FGL, and here are called ``non-c'' sources.




The positions, variability and spectral information given in the catalog provide an important starting point for the characterization of LAT unassociated sources. We can easily compare intrinsic properties of the 1FGL sources such as spectral index, curvature index and flux in different energy bands for both associated and unassociated populations, potentially providing insight into the likely classes of the unassociated sources.  


For the 1FGL catalog, the limiting flux for detecting a source with photon spectral index $\Gamma=2.2$ and Test Statistic of 25 \citep[TS = $2\Delta$log(likelihood)][]{Mattox1996} varied across the sky by about a factor of five (see Figure 19 of 1FGL). This non-uniform flux limit is due to the non-uniform Galactic diffuse background and non-uniform exposure (mostly arising from the passage of the \Fermi~observatory through the South Atlantic Anomaly).  

\begin{figure}[!ht]
\centering
\includegraphics[width=0.6\textwidth]{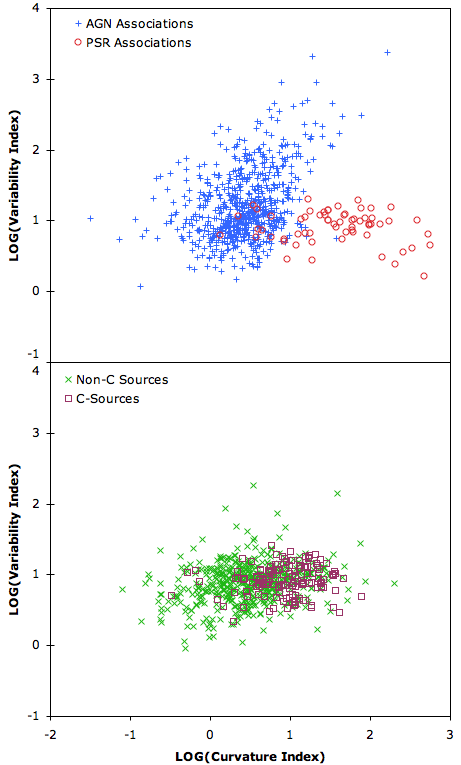}
\caption{Comparison of the 1FGL Variability Index versus Curvature Index 
for the associated sources (top panel) and unassociated sources (bottom). 
A separation between the AGN (crosses) and pulsar (circles) populations is 
evident. However the unassociated sources mainly lie in the region where those 
two populations overlap.
\label{fig:var_v_curve}}
\end{figure}

As discussed in 1FGL, when the variability and spectral curvature properties of \Fermi-LAT sources are compared against each other, a clear separation is visible between bright sources with AGN associations and those with pulsar associations. In Figure~\ref{fig:var_v_curve} (top panel), pulsars lie in the lower right-hand quadrant and AGN lie in the upper half. However, in the lower left-hand quadrant the two classes mix, making it difficult to distinguish between them. This region of parameter space is home to much of the unassociated source population (bottom panel). A closer look at these and other properties of the known sources gives clues to methods of separating the two major types, allowing us to classify some of the unassociated sources as likely members of one of these two source types (Section~\ref{sec:classification}).

\subsection{\label{sec:fluxdist}Source locations and Flux distributions}


The spatial distributions of the major source types (AGN, pulsars, unassociated sources) are given in Table~\ref{tbl:1FGLdistro}. It is clear that there is a significant excess of unassociated sources at low Galactic latitudes ($|b|<10\degr$) where 63\% of the detected sources have no formal counterparts, compared to only 36\% unassociated at $|b|>10\degr$. 

\begin{figure}[!ht]
\centering
\includegraphics[width=0.8\textwidth]{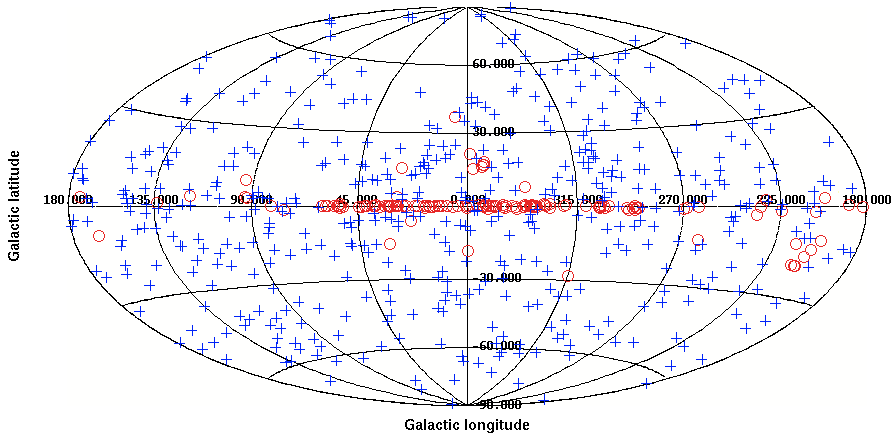}
\caption{1FGL sky map with the positions of the unassociated sources marked. 
Here, the non-c unassociated sources are indicated by crosses, the c-sources
by circles.
\label{fig:unassocskymap}}
\end{figure}

\begin{figure}[!ht]
\centering
\includegraphics[width=0.8\textwidth]{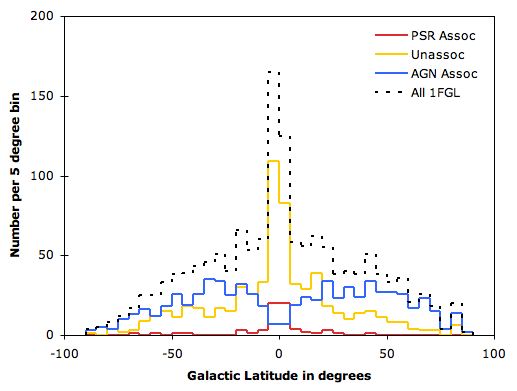}
\caption{Distribution of 1FGL sources types by Galactic latitude. The sources associated with AGN (blue line) show a clear deficit at low latitudes, while the same region hosts a large number of unassociated sources (yellow line) and identified pulsars (red line).
\label{fig:1FGLhisto}}
\end{figure}

Figure~\ref{fig:unassocskymap} shows the spatial distribution of LAT unassociated sources, with the positions of non-c sources shown as crosses and the c-source positions given by circles. As for the EGRET (3EG) catalog sources, the distribution is clearly not isotropic \citep{3rdCat}. One consideration when interpreting the distribution of unassociated 1FGL sources is that a number of the remaining unassociated sources are in low Galactic latitude regions where catalogs of AGNs have limited or no coverage, reducing the fraction of AGN associations. If we bin the different source types by Galactic latitude (Figure~\ref{fig:1FGLhisto}), we see a clear absence of AGN associations in the central $10\degr$ of the Galaxy ($|b|<5\degr$), while in the same region there is a spike in the number of unassociated sources.

The unassociated sources have an average flux of 3.1\,$\times$\,10$^{-9}$ \pflux ($E > 1$ GeV), while the associated population averages are 5.5\,$\times$\,10$^{-8}$ \pflux~for pulsars and 2.7\,$\times$\,10$^{-9}$ \pflux~for AGN. 

\begin{figure}[!ht]
\centering
\includegraphics[width=0.4\textwidth]{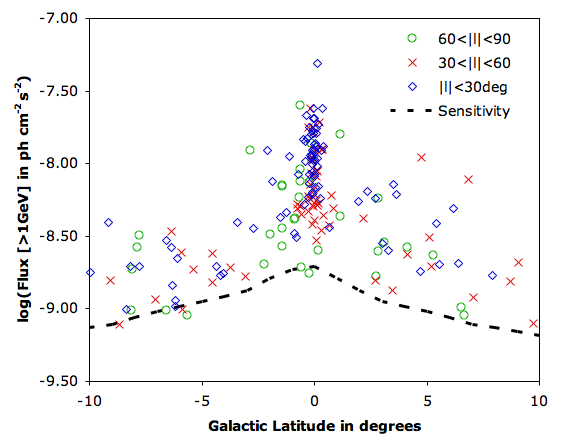}
\includegraphics[width=0.4\textwidth]{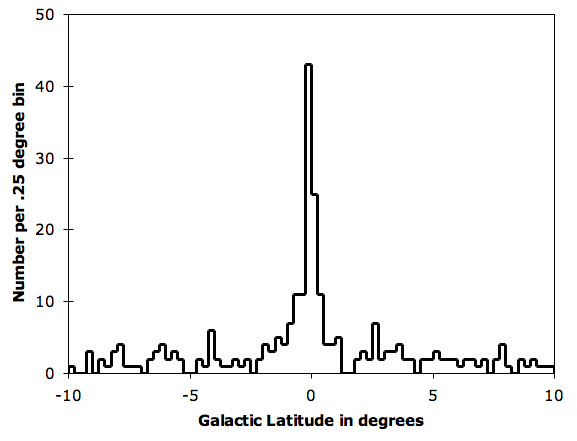}
\caption{Distribution of unassociated sources in the Galactic ridge. Left: Source flux ($E>1$ GeV) for all 1FGL sources as a function of Galactic latitude in three longitude bands. The dashed line shows the threshold flux for detectability of a source with a power-law spectrum of photon spectral index $\Gamma=2.2$ (from the 1FGL sensitivity map, at $|l|=0$). An increase in minimum flux is clearly visible for sources near $|b| = 0\degr$. Right: Unassociated source counts in $0\fdg25$ bins. A sharp peak in the number of unassociated sources is visible clustered along the central $0\fdg5$ of Galactic latitude.
\label{fig:latgaldist}}
\end{figure}

In the Galactic plane a $\gamma$-ray source must be brighter than at high latitudes in order to be detected above the strong Galactic diffuse emission. Figure~\ref{fig:latgaldist} (left panel) shows the 1FGL source flux distribution versus Galactic latitude for three longitude bands. It is clear that the Galactic plane ($|b| < 2\fdg5$) is dominated by Galactic diffuse emission, raising the flux detection threshold to $>$ 5\,$\times$\,10$^{-9}$ \pflux. This is reflected in the average flux of the unassociated c-sources which at 8.2\,$\times$\,10$^{-9}$ \pflux~is significantly higher than that for the non-c unassociated source population (1.7\,$\times$\,10$^{-9}$ \pflux~). Outside the central region of the Galaxy, the flux threshold is lower than that shown in Figure~\ref{fig:latgaldist}.


As was the case for COS-B and EGRET,  it is likely that a subset of the unassociated sources are spurious, resulting from an imperfect Galactic diffuse model. Such sources probably have very low significance, poor localization, and a spectral shape that mimics that of the Galactic diffuse emission itself. The c-sources in the 1FGL catalog are candidates to be sources of this type. 

As discussed in Section 4.7 of 1FGL, the latitude distribution of the Galactic ridge ($300^\circ < l < 60^\circ$) unassociated sources shows a sharp narrow peak in the central degree ($|b| < 0\fdg5$) of the Galaxy (Figure~\ref{fig:latgaldist}, right). If this feature is not an artifact, and we assume these sources originate in a Galactic population, then the scale height for this population must be $\sim$50 pc, to keep the average distance to the sources within the Galaxy. Such a scale height does not correspond to any known population of $\gamma$-ray sources, making it likely that a number of the sources in the Galactic ridge are spurious.

\subsection{\label{sec:spectralprops}Spectral properties}

\begin{figure}[!ht]
\centering
\includegraphics[width=0.6\textwidth]{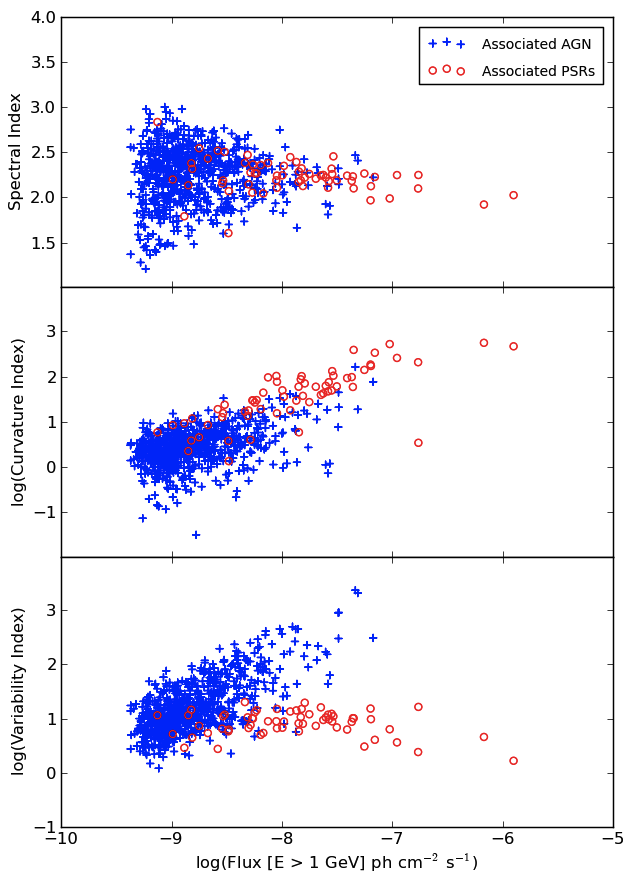}
\caption{Distributions with respect to flux of the spectral index (top), curvature index
(middle) and variability index (bottom) for the 1FGL associated and identified sources. It is clear
that the curvature index is dependent on source flux for both 
AGN (crosses) and Pulsar (circles) populations. The high flux pulsar with a low value for curvature index is the Crab pulsar.
\label{fig:idx_v_flux}}
\end{figure}

The 1FGL catalog provides spectral information that may be useful for distinguishing between different source classes. As part of the 1FGL analysis all sources were fit with a power-law spectral form and the spectral indices were included in the catalog. In addition, the catalog includes a ``curvature index'', which measures the deviation of the spectrum from the simple power-law form for each source. This means the curvature index is more a measure of the quality of the power-law spectral fit than of the intrinsic spectral shape. Figure~\ref{fig:idx_v_flux} shows the distributions of the spectral index (top panel) and curvature index (middle panel) with respect to flux. Neither of these parameters appears to discriminate well between the AGN and pulsar populations. In addition, the relationship is nearly linear for the curvature index, indicating that this parameter is strongly correlated with flux. That is, fainter sources have relatively poorly measured spectra that cannot be measured to be significantly different from power laws. This means that faint $\gamma$-ray sources provide less discriminating information than bright sources.

The majority of $\gamma$-ray AGN are blazars, which are relativistic jet sources with the jets directed toward the earth. An important property of blazars is their typical $\gamma$-ray spectral index, which offers some discrimination power between FSRQs (Flat Spectrum Radio Quasar) and BL Lacs \citep{LATlogNlogS}. The spectra of blazars in both of these sub-classes are typically well described as broken power laws in the LAT energy range, and the distributions of spectral indices for FSRQs and BL Lacs are compatible with Gaussians \citep{LBAS,1LAC}. However, because pulsar spectra are not well described by power-laws, the spectral index of a power-law fit is not a good discriminator between pulsar and AGN classes. 

As mentioned before, Figure~\ref{fig:idx_v_flux} (middle) shows that the curvature index for pulsars is strongly correlated with flux. This is primarily because many of the pulsars detected in the 1FGL catalog are strong $\gamma$-ray sources, with brighter pulsars having a more significant spectral curvature than fainter pulsars. Unfortunately, the broken power-law spectral forms of bright blazars \citep[e.g.,][]{LAT3C454} also have the effect of inducing a correlation between curvature index and flux for LAT blazars. 

There are few sources with significant detections in all of the five spectral bands used to calculate the curvature index. Only 36 of the 630 unassociated sources are strongly-enough detected to have flux measurements reported in each band in the 1FGL catalog, as this requires a TS $>$ 10 in each energy range. By contrast, 181 of the unassociated sources are detected in only a single band, with an additional 88 sources having upper limits in all the spectral bands (i.e., are only detected when data at all energies are combined).

\subsection{\label{sec:varprops}Variability properties} 

In the 1FGL catalog the variability index for each source was defined as the $\chi^{2}$ of the deviations of eleven monthly (30-day) source flux measurements from the average source flux \citep{1FGL}. While this value increases with flux for AGN, it does not do so for the pulsars (Figure~\ref{fig:idx_v_flux}, bottom panel), making variability a much better discriminator between the two major classes. One property of blazars is that they are frequently significantly variable in $\gamma$ rays \citep{LBAS:variability}. Their fluxes can vary up to a factor of five on time scales of a few hours and by a factor of 50 or more over several months. As a consequence, their characteristic variability can serve as a primary discriminator. This property has been used to turn some \Fermi-LAT AGN associations into identifications due to their timing properties \citep{LBAS:variability,planeblazar}. For variability to be a useful indicator, the time scale must be adapted to the source significance. Indeed, for a faint source, the variability needs to be tested on longer time scales than for a bright source. All sources in the 1FGL catalog were processed in the same way, regardless of flux. Thus, for the many 1FGL sources not bright enough to be significantly detected on monthly time scales the 1FGL variability index is not a sensitive discriminator of variability.

Pulsars, on the other hand, are generally steady sources. Where variability has been seen in $\gamma$ rays, it has been attributed to flares in the nebular contribution of a PWN, rather than to the pulsar itself \citep{Tavani2011,LATCrabvar}. This flux stability places pulsars in extreme opposition to AGN in the $\gamma$-ray regime. Essentially any significant detection of variability in an unassociated source is enough to make a pulsar classification extremely unlikely.

In the 1FGL catalog, 241 sources were found to be variable at a formal confidence level of 99\% (variability index $>$ 23.21). Of these, 2 are HMXBs, 221 are AGN, and 18 are unassociated. Variability in bright sources is easier to detect as the source flux is typically above the sensitivity threshold in each monthly bin. For the lower-flux unassociated sources, however, we need a method to improve the detection of variability. Using the fractional variability (discussed in Section~\ref{sec:discriminators}, is one such method.

\subsection{\label{sec:lognlogs}Comparison with source modeling}

We can also examine what source distributions we might expect given the populations of source types in 1FGL.  We do this by first estimating the total number of detected AGN in 1FGL, which we derive from a model population. To quantify the total number of AGN, we model the population and then apply that model to the 1FGL catalog. 

\begin{figure}[!ht]
\centering
\includegraphics[width=0.8\textwidth]{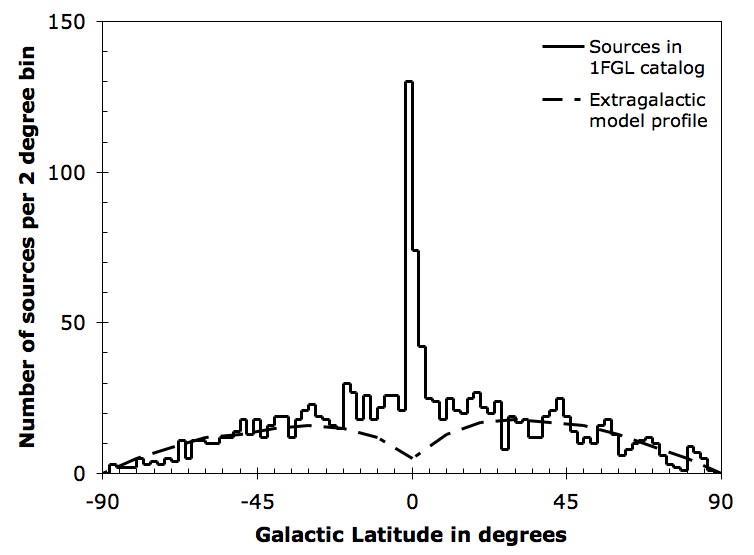}
\caption{Latitude profile of the 1FGL sources and the extragalactic source model profile (dashed line)
\label{fig:profile}}
\end{figure}

We use the \Fermi-LAT Log{\it N}-Log{\it S} distribution (the distribution of the number {\it N} of sources detectable at a given sensitivity {\it S}) for AGN \citep{LATlogNlogS}, and the 1FGL sensitivity map \citep[Figure 19 of][]{1FGL} to generate a Galactic map that contains the number of the expected AGN at each position in the sky. Summing these results over Galactic longitude we obtain the AGN latitude profile shown in Figure~\ref{fig:profile}. Integrating the AGN model allows us to estimate the number of expected AGN in the sky. By subtracting the number of AGN found by the model from the sources in the 1FGL catalog we obtain the number of Galactic sources in and out of the plane. This gives the Galactic source estimates for the unassociated source list. 

Table~\ref{tbl:sourcedistribution} compares the 1FGL source counts with this model for low and high-latitude regions. It is clear that the group of sources that is most difficult to associate are those of Galactic origin at low Galactic latitudes. This is likely due to the presence of a population of spurious sources in that region in the 1FGL catalog.

At high Galactic latitudes, pulsars are the second most numerous class of identified $\gamma$-ray sources; most of those are millisecond pulsars (MSPs). From the set of 1FGL pulsars and the new pulsar associations discussed herein, we find that more than a third of the $\gamma$-ray pulsars known to date are MSPs. If we then assume that $\sim$50\% of the 271 unassociated sources that are expected to be Galactic sources are pulsars (based on the fraction of Galactic sources which are pulsars as given in Table~\ref{tbl:1FGLdistro}), and one third of those pulsars are MSPs, we find that we expect 45 new MSPs in the 1FGL unassociated source list. Of the 31 new MSPs discovered to date (Section~\ref{sec:Radio_psr}), 28 are at high Galactic latitudes, suggesting that an expected number of 45 is not unreasonable.

At low Galactic latitudes the source content is more diverse, with half the sources being compact objects (pulsars and X-ray binaries) and nearly half being extended sources (SNRs and PWNe). If the unassociated sources have a similar distribution, then there will be $\sim$100 pulsars and $\sim$100 SNRs/PWNe.

\section{\label{sec:classification}Classification of unassociated sources}

The spatial, spectral and variability properties discussed in Section~\ref{sec:sample} provide a framework that allows us to try to predict the expected source classes for the sources that remain unassociated. This is done by using the properties of the associated sources to define a model that describes the distributions and correlations between measured properties of the $\gamma$-ray behavior of each source class. This model is then compared to the $\gamma$-ray properties of each unassociated source. Generating the model requires an associated source parent population with enough members to describe the behavior well. For this reason, we have focused only on AGN and pulsars as the input source populations.

To create a model, it is necessary to use $\gamma$-ray properties that are clearly different between the parent populations. For 1FGL, the best parameters are the spectral index, curvature index, and variability index, as well as hardness ratios between the different spectral bands. In addition, it is important that the properties used to generate the model not be related to source significance, as this will bias the results. However, as discussed in Section~\ref{sec:fluxdist}, the curvature index appears to be dependent on the source flux, and thus is not a good indicator of the parent population. Also, the spectral index and hardness ratios are both spectral indicators, so they overlap in functionality. Since the hardness ratios provide more information about spectral shape than the spectral index, they are preferred for this analysis. 

To generate valid classifications, we must first define new parameters that allow intrinsic properties to be compared rather than relative fluxes. With the new parameters in hand, we can generate classification predictions using multiple methods, and compare these predictions to each other. However, the results from these techniques assume that the training samples and test samples have the same distributions of intrinsic properties. This is not true for the 1FGL unassociated sources, as they are more frequently found in the plane of the Galaxy with elevated background levels and in confused regions. To help compensate for this difference, we will validate the results against an independent set of classified sources.

\subsection{\label{sec:discriminators}Improving source type discriminators}

To mitigate the effect of low fluxes on the determination of the band spectra, it is necessary to define additional comparative parameters that remove the significance dependency. In this case, the 1FGL catalog provides a set of fluxes in five bands for each source from which we can find hardness ratios. To get a normalized quantity the hardness ratios are constructed as:
\begin{equation}
\centering
HR_{ij} = (EnergyFlux_{j} - EnergyFlux_{i}) / (EnergyFlux_{j} + EnergyFlux_{i})
\label{HR_equation} 
\end{equation} 
This quantity will always be between $-$1 and 1; $-$1 for a very soft source ([high]EnergyFlux$_{j}$ = 0) and +1 for a very hard source ([low]EnergyFlux$_{i}$ = 0). Here energy flux in log(E) units (i.e. $\nu$F$_\nu$) is used instead of photon flux because the definition works well only when the quantities are of the same order. This is true for the energy fluxes (because the spectra are not too far from a E$^{-2}$ power law) but not for photon fluxes. 

It is also possible to define a quantity that discriminates curvature by combining two hardness ratios, preferably from bands with a high number of detected sources. Here we use (HR$_{23}$ - HR$_{34}$), where bands 2, 3, and 4 are for 0.3$-$1, 1$-$3, and 3$-$10 GeV respectively. This hardness ratio difference, or curvature, is positive for spectra curved downwards in $\nu$F$_\nu$ (like pulsars), zero for power laws and negative for spectra curved upwards (with a strong high-energy component). 

\begin{figure}[!ht]
\centering
\includegraphics[width=0.6\textwidth]{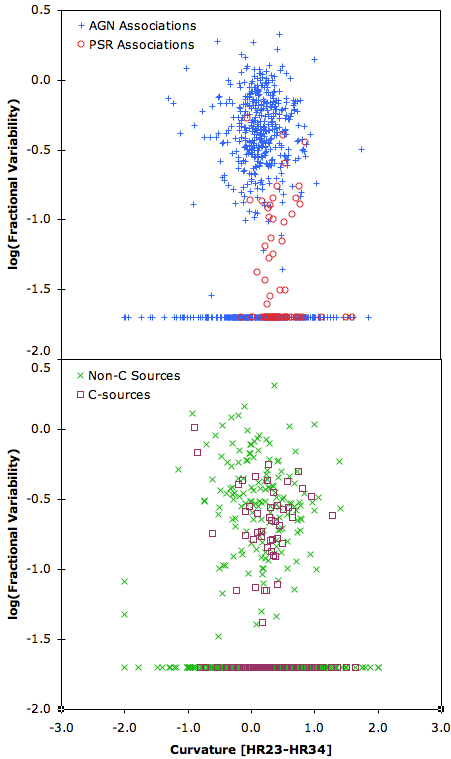}
\caption{The fractional variability vs. hardness ratio difference. Top: the 1FGL associated AGN (blue crosses) and pulsars (red circles). Bottom: the 1FGL non-c unassociated sources (green crosses) and the c-sources (purple squares).
\label{fig:Frac_v_HRdiff}}
\end{figure}

To remove the source significance dependency for variability, we use the fractional variability \citep[as defined in Equation 5 of][]{1FGL} instead of the variability index. The fractional variability is:
\begin{equation}
\centering
FracVar = \sqrt{\frac{\sum_i{(Flux_{i} - Flux_{\rm av})^{2}}}
				{(N_{\rm int}-1)Flux_{\rm av}^{2}} - 
			\frac{\sum_i{\sigma_{i}^{2}}}
				{N_{\rm int} Flux_{\rm av}^{2}} -
				f_{\rm rel}}
\label{FV_equation} 
\end{equation} 
where $N_{int}$ is the number of time intervals (11 in 1FGL), $\sigma_{i}$ is the statistical uncertainty in $F_i$, and $f_{\rm rel}$ is an estimate of the systematic uncertainty on the flux for each interval. (Here we use 3\% as in 1FGL.) For some 1FGL sources the quantity inside the square root is negative. Those sources are assigned a fractional variability of 2\%. Figure~\ref{fig:Frac_v_HRdiff} shows fractional variability versus the curvature (HR$_{23}$ - HR$_{34}$) for both the associated (top panel) and unassociated (bottom panel) 1FGL sources. 

To allow the largest possible sample when performing the classification, we use the actual best fit values in calculating both these quantities, even when the 1FGL analysis reported only 2-sigma upper limits. For such sources the variable is not well constrained. However, it will contribute to the distribution used to determine the classifications and may provide a small amount of additional discriminating power.



\subsection{\label{sec:ClassTree}Classification using Classification Trees}

We have implemented two different data mining techniques to determine likely source classifications for the 1FGL unassociated sources: Classification Trees (this section) and Logistic Regression (Section~\ref{sec:LogRegress}). Both techniques use identified objects to build up a classification analysis which provides the probability for an unidentified source to belong to a given class. We applied these techniques to the sources in 1FGL to provide a set of classification
probabilities for each unidentified source.

Classification Trees are a well-established class of algorithms in the general framework of data mining and machine learning \citep{CT_book}. The general principle of machine learning is to train an algorithm to predict membership of cases or objects to the classes of a dependent variable from their measurements on one or more input variables. The advantage of this class of algorithms is the ability to produce a unique predictor parameter that takes into account several input quantities simultaneously. Other well-known flavors of machine learning algorithms include artificial neural networks, support vector machines and bayesian networks.

The purpose of analyses via tree-building algorithms is to determine a set of {\it if-then} logical conditions (called tree-nodes) that permit accurate prediction or classification of cases: the training procedure generates a tree in which each decision node contains a test on some input variable's value. The trees in this analysis are built through a process known as binary recursive partitioning, which is an iterative process of splitting the data into partitions. Initially all the records in the training set are assigned to a single class: the algorithm then tries breaking up the data, using every possible binary split on every field. The peculiar advantage of Classification Trees for our specific application is their flexibility in handling sparse or uneven distributions.

The specific flavor of algorithm used in this case was Adaptive Boosting, which reweights the importance of input sources at each step of the classification. The training of boosted decision trees is a recursive procedure, whereupon the weights of each incorrectly classified example are increased at each step, so that the new classifier focuses more on those examples. The output of such a procedure is actually a collection of trees, all grown from the same input sample: the selection will be done on a majority vote on the result of several decision trees (200 in the present case). In the following text, we will always refer to a single Classification Tree for simplicity, even though the real classifier is a much more complex object.

The training and application of Classification Trees to this specific analysis has been performed using TMine, an interactive software tool for processing complex classification analyses developed within the Fermi-LAT collaboration \citep{Alex_tmine}. TMine is based on ROOT \citep{root_paper}, a very popular data analysis framework for high energy physics experiments. For the processing and parallel evaluation of multivariate classification algorithms, TMine utilizes the ROOT Toolkit for Multivariate Analysis (TMVA) \citep{tmva_paper}.

\subsubsection{Selection of the relevant training variables}

The first step of the CT analysis is to select a sample of data to build the predictor variable. We decided to focus on the two most abundant classes of objects in the \Fermi~catalog, AGN and pulsars, and to train a Classification Tree to discriminate between them. We trained a predictor using all the associated AGNs and pulsars in the 1FGL catalog \citep{1FGL}. Because of the spectral similarities with pulsars, potential associations for sources near SNRs have been compounded with the pulsars sample. The output of this training process is a parameter (the predictor) which describes the probability for a new source to be an AGN.

Choosing the most appropriate set of variables for training the Classification Tree is a very delicate step in the analysis. It is extremely important to ensure that the selected variables are not dependent on the flux, the location or the significance of the source: this check can be accomplished by comparing the distribution of the various parameters for associated and unassociated sources. Physical considerations about the $\gamma$-ray properties of each source class also guided us in the choice of the most effective variables for discriminating AGN from pulsars (Section~\ref{sec:discriminators}).

After exploring most of the parameters in the 1FGL catalog, we selected a set of variables that includes: the curvature (HR$_{23}$ - HR$_{34}$), the spectral index and the fractional variability of each source, plus the Hardness Ratios for the 5 energy bands in the catalog. Table~\ref{CT_ranking} ranks the relative importance of the different variables at distinguishing AGNs from pulsars: the weight of each variable was computed by the Classification Tree algorithm during the training process.

As described in section~\ref{sec:discriminators}, we used for training the actual best fit values for each variable, even when the 1FGL analysis reported only upper limits. In case of faint sources that are not detected in one of the energy bands, some of the Hardness Ratios will be very close to -1 or 1: this is when the ability of Classification Trees in handling sparse distributions is particularly useful. Similar considerations apply to the fractional variability, which is very close to zero for pulsars, but varies for AGNs.

We chose not to use the Galactic latitude as input to the CT in order to avoid biasing our selection against AGN situated along the Galactic plane and pulsars (especially MSPs) situated at high Galactic latitude. Furthermore, this choice gives us the opportunity to use the latitude distributions of the different populations as a cross check of our result. (The new pulsar candidates should be mainly distributed along the Galactic plane, while the AGN candidates should be isotropically distributed).

While training the Classification Tree, 30\% of known AGNs and pulsars, randomly selected from the input sample, were kept aside for cross-validation of the method. Such cross-validation was performed by comparing the predictor distributions for the training and testing samples via a Kolmogorov-Smirnov test. The result of the test provides a 93\% probability for the AGN distributions and a 47\% probability for the pulsar distributions, which make the test fully satisfactory.

It must also be noted that the sources associated with a different class than AGN or pulsar have been excluded from this training procedure, for a total of 24 sources. We cannot treat these 24 sources uniformly as ``background'', because of the smallness of their sample and the diversity of their spectral properties. However, it is possible to estimate the contamination to the candidate AGN and pulsar samples deriving from the likely presence of these "other" sources in the unassociated sample.

\subsubsection{Output of the Classification Procedure}

The second step of the analysis consists of deriving the predictor variable for unassociated sources by applying the Classification Tree that was trained in the previous step. The resulting predictor is a parameter that describes the probability that each of the unassociated sources is an AGN. The predictor is included in Table~\ref{tbl:gamma_table} which lists all 630 unassociated \Fermi-LAT 1FGL sources, and combines results for all the analyses discussed within this paper.

\begin{figure}[!ht]
\centering
\plottwo{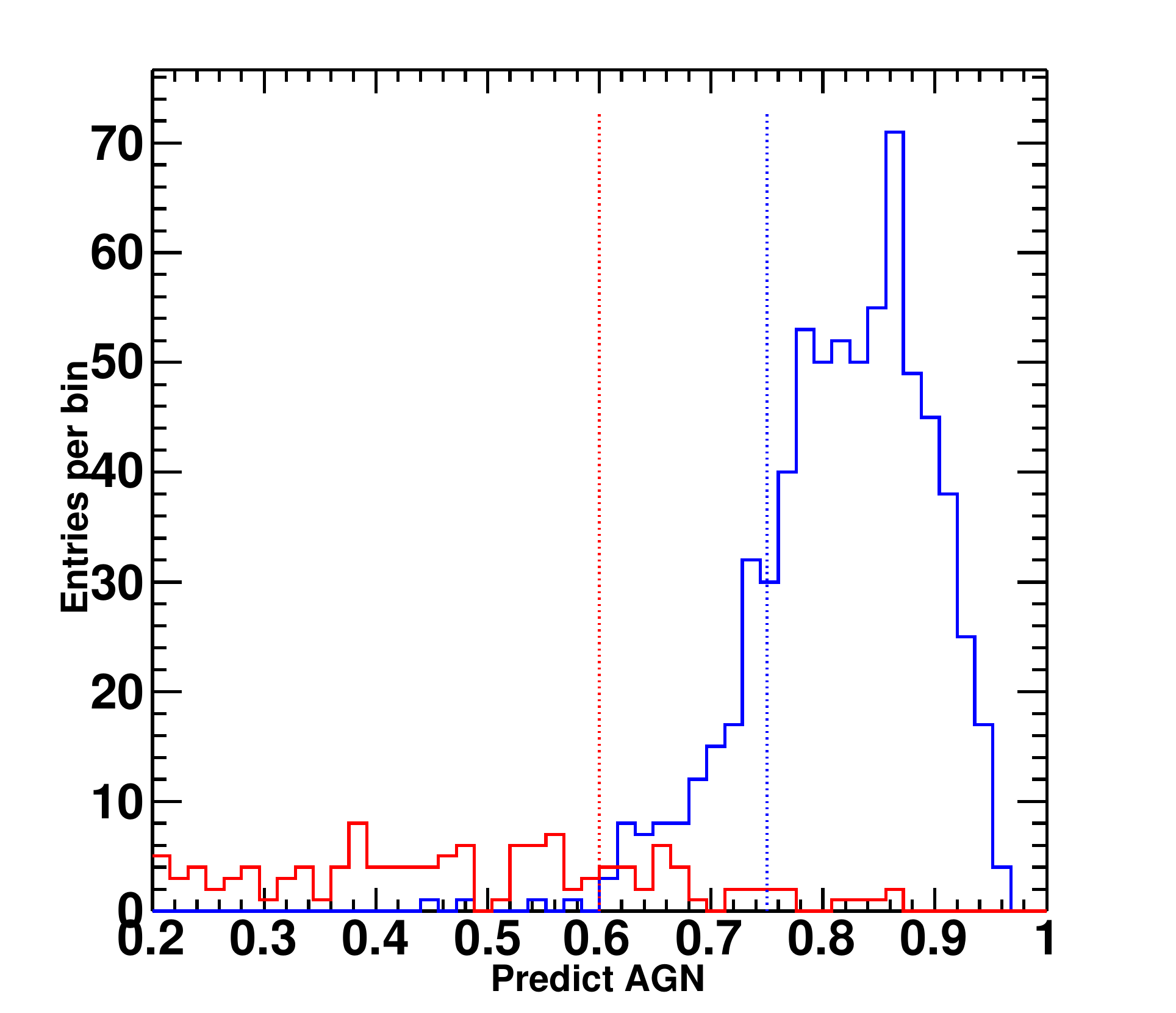}{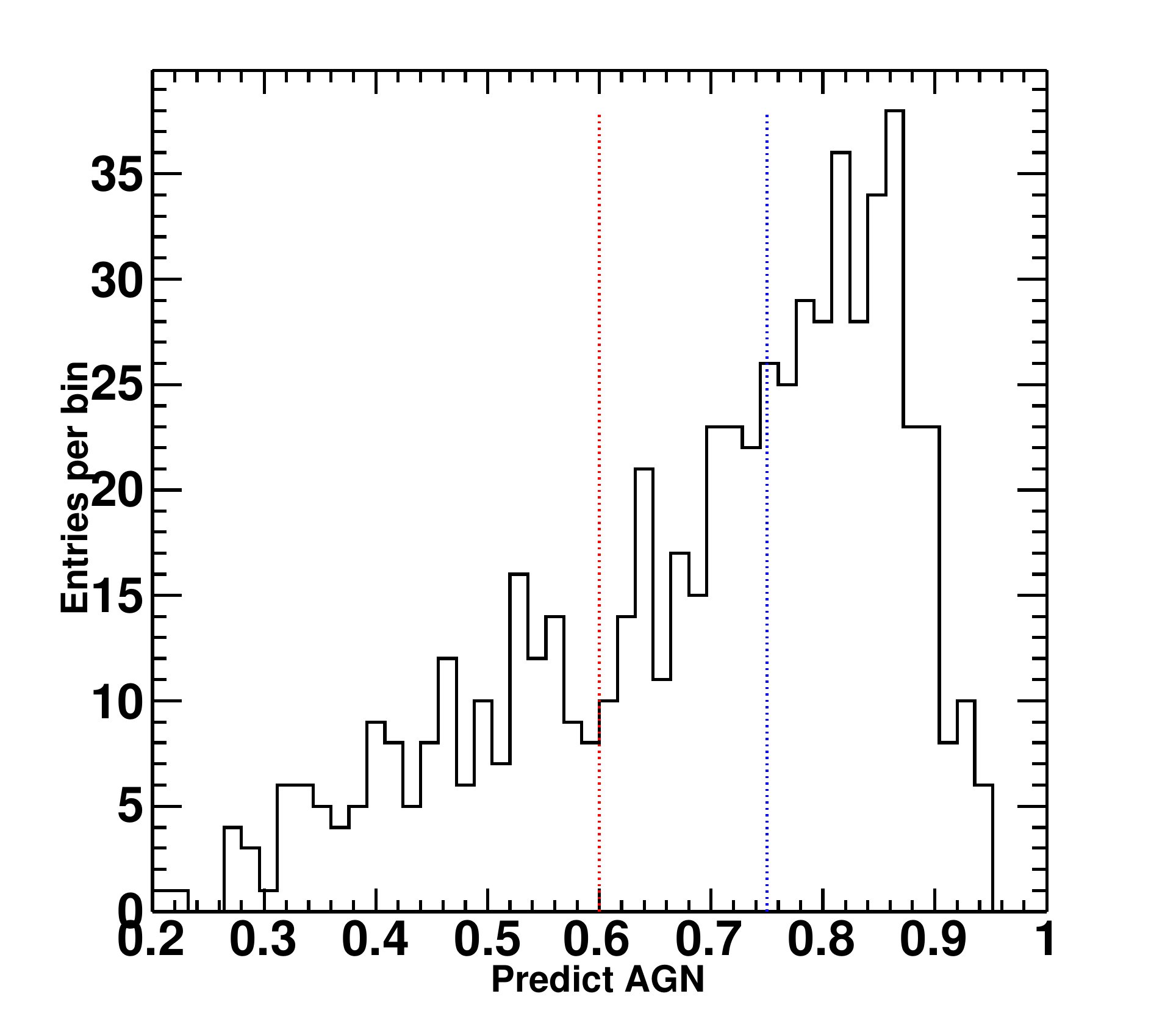}
\caption{Distribution of the Classification Tree predictor. Vertical lines indicate the value of the thresholds we set to identify AGN candidates (Predictor $>$0.75) and pulsar candidates (Predictor $<$0.6). Left: sources of the 1FGL catalog identified as pulsar (red) and AGN (blue). Right: Distribution of the predictor for unassociated sources. 
\label{fig:CT_distros}}
\end{figure}

Figure~\ref{fig:CT_distros} shows the distribution of the predictor for the 1FGL associated sources used in the training of the tree (left panel) and the distribution of the predictor for the unassociated sources (right panel). The global shapes of the two distributions are clearly different, with an apparent excess of pulsar-like sources among the unassociated sources when compared to the associated source distribution. This may be due to the presumably different fractions of AGN and pulsars in the associated and unassociated samples, or there may be an additional contributing component.  Nevertheless, the distribution of associated sources clearly shows that we can select a set of AGN and pulsar candidates with high confidence, when choosing the appropriate fiducial regions.

We set two fiducial thresholds: all the sources with a predictor greater than 0.75 are classified as AGN candidates while all the sources with a predictor smaller than 0.6 are classified as pulsar candidates. All the sources with an intermediate value of the predictor remain unclassified after the CT analysis. The choice of these boundaries is optimized for an efficiency of 80\% for the two source classes in order to keep the misclassification fraction under 2\% (the misclassified fraction for a certain efficiency is determined by the width of the predictor distribution). Here, 80\% of AGN associations in 1FGL have a predictor greater than 0.75 and 80\% of pulsars have a predictor smaller than 0.6. 

In this case, the extrapolation from the value of the predictor to the probability of class membership was performed empirically from the combined input sample (which includes both the training and testing samples). The expected misclassification fraction in each class was also evaluated with the same method. This analysis was repeated using the training and testing samples separately and yielded identical results. A more complex study used the area under the Receiver-Operating-Characteristic (ROC) curve that is obtained by plotting all combinations of true positives and the proportion of false negatives generated by varying the decision threshold. This study provided similar extrapolation results, but more optimistic misclassification fractions: we therefore decided to rely on the more conservative misclassification estimation provided by the combined input sample.

The predictor distribution for the 24 sources that were not used during the training procedure can be used to estimate the further contamination from these sources to the AGN and pulsar candidate distributions: we expect that up to 2\% of the newly classified AGN candidates and up to 4\% of the newly classified pulsar candidates will indeed belong to one of the "other" classes (galaxies, globular clusters, supernova remnants, etc.).


\subsection{\label{sec:LogRegress}Classification using Logistic Regression}

Another approach to assign likely classifications for the 1FGL unassociated sources is the Logistic Regression (LR) analysis method \citep{LR2000}. Unlike the CT analysis, LR allows us to quantify the probability of correct classification based on fitting a model form to the data.

LR is part of a class of generalized linear models and is one of the simplest data mining techniques. LR forms a multivariate relation between a dependent variable that can only take values from 0 to 1 and several independent variables. 
When the dependent variable has only two possible assignment categories, the simplicity of the LR method can be a benefit over other discriminant analyses.

In our case, the dependent variable is a binary variable that represents the classification of given 1FGL unassociated source.  Quantitatively, the relationship between the classification and its dependence on several variables can be expressed as:
\begin{equation}
 P=\frac{1}{\left( 1+e^{-z}\right) }
\end{equation}
where $P$ is the probability of the classification, and $z$ can be defined as a linear combination:
\begin{equation}
 z=b_0+b_1x_1+b_2x_2+...+b_nx_n
\end{equation}
where $b_0$ is the intercept of the model, the $b_i$ (i = 0, 1, 2, ..., n) are the slope coefficients of the LR model and the $x_i$ (i = 0, 1, 2, ..., n) are the independent variables. Therefore, LR evaluates the probability of association with a particular class of sources as a function of the independent variables (e.g. spectral shape or variability).

Much like linear regression, LR finds a ``best fitting'' equation. However, the principles on which it is based are rather different. Instead of using a least-squared deviations criterion for the best fit, it uses a maximum-likelihood method, which maximizes the probability of matching the associations in the training sample by optimizing the regression coefficients. As a result, the goodness of fit and overall significance statistics used in LR are different from those used in linear regression.

\subsubsection{Selection of the training sample and the predictor variables}

As LR is a supervised data mining technique, it must be trained on known objects in order to predict the membership of a new object to a given class on the basis of its observables. As with CT analysis, we trained the predictor using the pulsar and AGN associated sources in the 1FGL catalog \citep{1FGL}. Like the CT analysis, the output of this training process is the probability that an unidentified source has characteristics more similar to an AGN than to a pulsar.

To evaluate the best predictor variables for the LR analysis, we used the likelihood ratio test, comparing the likelihoods of the models not including (null hypothesis) and including (alternative hypothesis) the predictor variable under examination. We started by using the fractional variability, the spectral index, the hardness ratios for the 5 energy bands in the catalog and the position on the sky (i.e. the Galactic latitude and longitude). The value of the likelihood ratio test is the p-value, and is useful in determining if a predictor variable is significant in distinguishing an AGN from a pulsar. If the p-value for a given predictor variable is smaller than the significance threshold $\alpha$ (0.05) then the predictor variable is included in the multivariate LR model. We did not include the curvature value ($HR_{23}-HR_{34}$) in this evaluation because the LR analysis does not work well with predictor variables that are linearly dependent on other predictor values. 

We then calculated the significance of each predictor variable to find the resulting LR coefficients. The list of the LR predictor variables with the relative values of the maximum likelihood ratios can be found in Table~\ref{tbl:LRpredicts}. 

While AGN are isotropically distributed and pulsars are concentrated along the Galactic plane, we wanted to see whether our multivariate LR model was able to recognize this effect. The results indicate that Galactic latitude and longitude are not significant at the $\alpha=0.05$ (5\% significance) level. Moreover we find that also $HR_{12}$ is not highly significant in the LR analysis. It is interesting to note that $HR_{12}$, in the univariate LR model, is quite significant (p-value=0.02) to distinguish between AGNs and pulsars but in a multivariate LR analysis it loses its significance. In Table~\ref{tbl:LRpredicts} those predictor variables selected for the LR model are above the line and those we did not select lie below the line.

\subsubsection{Defining thresholds}

Next we derive the predictor variable for 1FGL unassociated sources by applying the trained classification analysis to those sources. Since the LR analysis used AGNs as primary source type, the output parameter ($A$) listed in Table~\ref{tbl:gamma_table} describes the probability that an unassociated source is an AGN. The probability that an unassociated source is a pulsar is $P=1-A$ (because we are modeling the behavior of AGNs as ``opposite'' of the behavior of the pulsars based on the predictor variables). 

In principle, the dependent variable is a binary variable that represents the presence or absence of a particular class of objects. We could have selected ``pulsars'' and ``non-pulsars'' (e.g. all other 1FGL associated sources) to teach the model to recognize the new pulsars, and done similarly for the AGNs. We did not follow this approach because there are no source populations in 1FGL other than AGNs and pulsars with sufficient numbers to significantly affect the results. By focusing on ``opposing'' characteristics, we improve the efficiency of classifying new AGN or pulsar candidates.

As with the CT analysis we defined two threshold values, one to classify an AGN candidate ($C_A$) and one to classify a pulsar candidate ($C_P$). We chose these two thresholds by analyzing the ROC curves so that 80\% of the AGN associations in 1FGL would have a predictor value greater than $C_A$ and 80\% of the pulsars would have a predictor value smaller than $C_P$, and to result in very low contamination. Using this principle we set $C_A$ to 0.98 and $C_P$ to 0.62. With these thresholds, only 1\% of AGNs are misclassified as pulsars, while 3\% of pulsars are classified as AGN. 

To estimate how accurately our predictive model performs in practice, we cross-validated using only  the 756 pulsars and AGNs in the 1FGL catalog. We held out 75 sources to be the testing data set, and we used the remaining 681 for training. We repeated this procedure 10 times, using a different set of 75 test sources in each data set. At the end, this 10-fold cross-validation showed that the average testing efficiency rates for these threshold values are $75\%$ for pulsars and $80\%$ for AGNs, and that the average testing error rates (false positives) are very low, $0.05\%$ for pulsars and $0.02\%$ for AGNs. The 5\% lower success rate for the pulsars is likely due to low statistics in the test samples.

\begin{figure}[!ht]
\centering
\plottwo{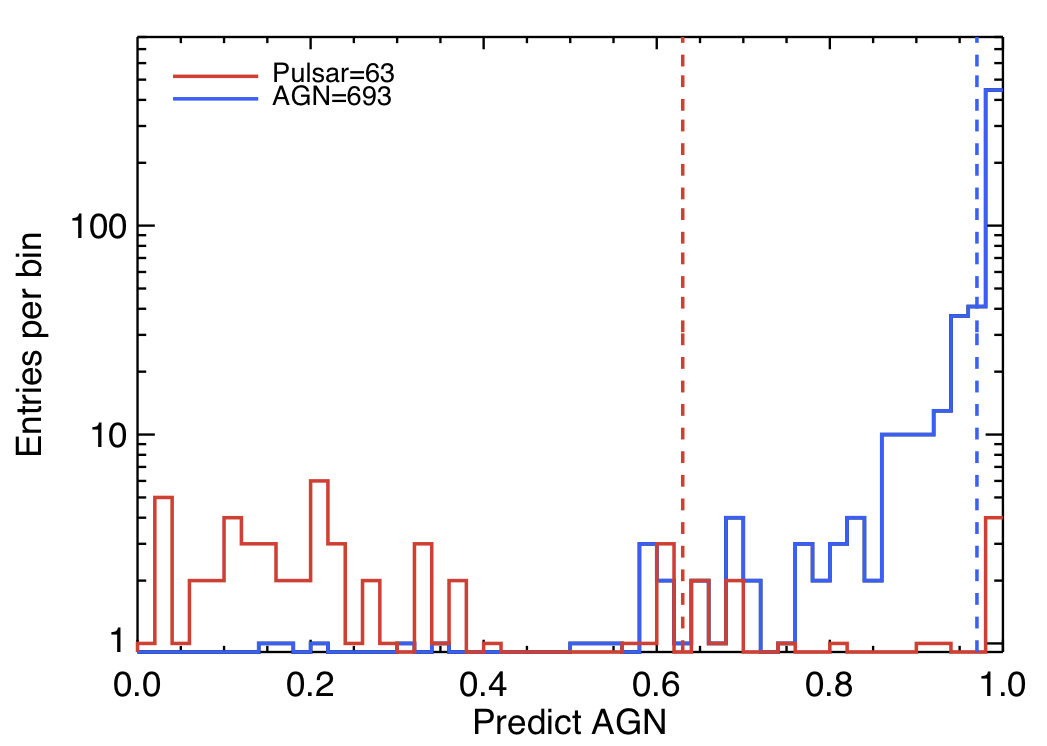}{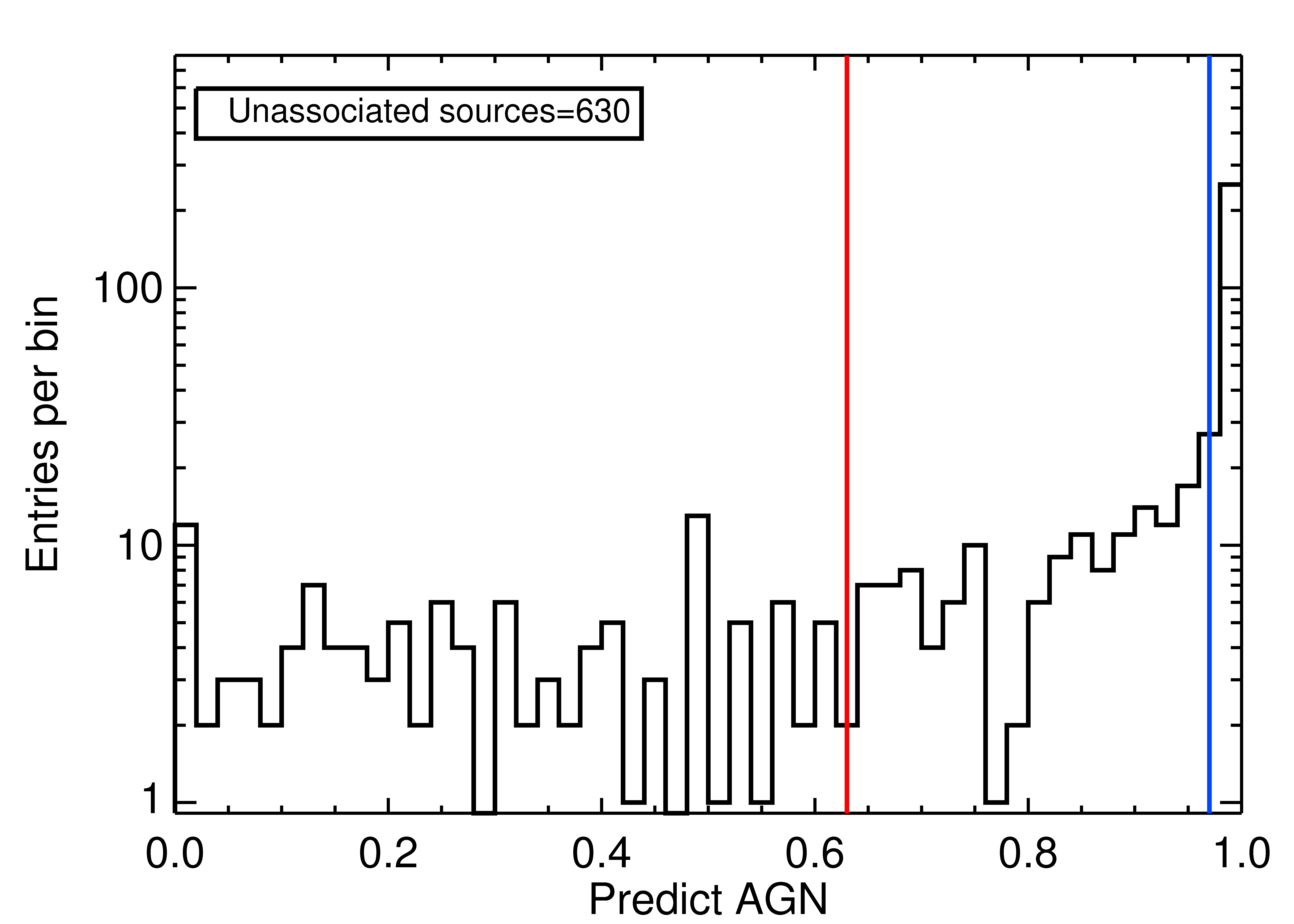}
\caption{Distribution of the Logistic Regression predictor. Vertical lines indicate the value of the thresholds we set to identify pulsar candidates (Predictor $<$0.62) and AGN candidates (Predictor $>$0.98). Left: sources of the 1FGL catalog identified as pulsars (red) and AGNs (blue). Right: for 1FGL unassociated sources. 
\label{fig:Figure23}}
\end{figure}

If we apply the model to the 1FGL unassociated sources we find that 368 are classified as AGN candidates ($P>0.98$), 122 are classified as pulsar candidates ($P<0.62$) and 140 remain unclassified after the LR analysis. The distributions of 1FGL associated and unassociated sources as a function of the probability of being pulsars are shown in the Figure~\ref{fig:Figure23}. The thresholds for assigning pulsar candidates and AGN candidates are indicated in the figure. It is important to note that in order to meet the acceptance threshold of 80\% of the known pulsars, we are including a large range of predictor values with very few pulsars. This may result in over-predicting the number of pulsars in unassociated sources.



\subsection{\label{sec:CombClass}Combining the two classification methods}

The two classification techniques gave somewhat different results. Of the 630 unassociated sources in 1FGL, both techniques agreed on the appropriate classification for 57.6\% of the sources (363), while they gave conflicting classifications for 5.4\% (34 sources). The remaining 253 sources were left unclassified by one or both techniques (see Table~\ref{tbl:Class_summary}). Studies comparing these classification techniques \citep{CTvLRcomp} have indicated that in data sets with good separability between the discriminating characteristics, the CT analysis should provide a more robust result. However, it is evident from the right panels of Figures~\ref{fig:CT_distros}~and~\ref{fig:Figure23} that the signal to noise for the unassociated sources does not provide such clear separability.

Since the purpose of this analysis is to provide candidate sources for follow-up multiwavelength studies, we use the positive results from both techniques to generate our candidate lists.
From these two methods we can synthesize a final set of classifications for each source, where:

\begin{itemize}
\item AGN candidates must be classified by at least one method, and the other method must not disagree (that is, not classify it as a pulsar).
\item Pulsar candidates must be classified by at least one method, and the other method must not disagree (that is, not classify it as an AGN).
\item Unclassified sources are not classified by either method.
\item ``Conflicting" sources are those that have been assigned opposite classifications (one AGN and one pulsar) by the two different methods
\end{itemize}

\begin{figure}[!ht]
\centering
\includegraphics[width=0.6\textwidth]{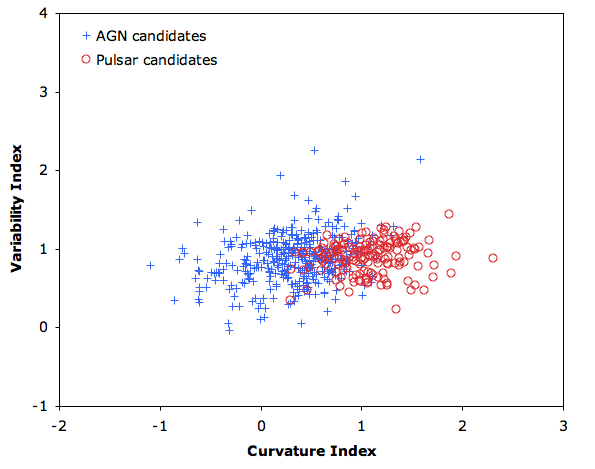}
\caption{Variability index versus curvature index for 1FGL unassociated sources classified as AGN (blue crosses) and pulsar candidates (red circles).
\label{fig:Class_scatter}}
\end{figure}

Based on these definitions, there are 396 AGN candidates (269 are classified as AGN by both methods), 159 pulsar candidates (72 classified as pulsars by both methods), 41 unclassified sources, and 34 conflicting sources in the 1FGL unassociated source list. Figure~\ref{fig:Class_scatter} shows the curvature-variability distribution of the newly classified AGN and pulsar candidates based on this synthesis of the two methods. We see that the unassociated sources have been separated into two populations with some overlap between them. Comparing this distribution to Figure~\ref{fig:var_v_curve} (top panel), we see that this separation follows the separation seen between the associated AGN and pulsars.

\section{\label{sec:newassoc}Recent association efforts}

In order to validate the results of the classification methods, we must obtain an independent set of associations from those used to train the two methods. For this, we look to association efforts that have taken place since the release of the 1FGL catalog. This section will present the new associations, while the classification validation is reported in Section~\ref{sec:discuss}.

The associations listed in 1FGL were based on likely $\gamma$-ray-producing objects, i.e. those with energetic, non-thermal emission. The first LAT catalog of Active Galactic Nuclei (AGN) published shortly afterward found high-confidence AGN associations for 671 high Galactic latitude 1FGL sources, with an additional 155 LAT sources included in the low-latitude and lower confidence association lists \citep[1LAC;][]{1LAC}.  The 1LAC association method was the same as for 1FGL, but the acceptance threshold for association was lower than for 1FGL.

For the unassociated sources in this paper, which by definition do not have plausible counterparts among the candidate catalogs used for comparison with the LAT sources, we must look beyond the obvious candidate source classes.  Even with the improved source locations provided by the \Fermi-LAT, however, positional coincidence with a particular object is insufficient to claim an association.  

If potential candidates can be found, then additional tests, based on spatial morphology, correlated variability, or physical modeling of multiwavelength properties, offer the opportunity to expand the list of associations. X-ray, optical, or radio candidate counterparts all have better localizations than the $\gamma$-ray sources, so that a candidate in one of these wavelength bands can be matched with those in others. Also, most (if not all) of the catalogs and observations used here to find new associations are not complete surveys of the sky. Therefore the lack of an association for a 1FGL source does not mean that the source cannot be associated. In this section we present only a preliminary report of the results from the many ongoing efforts to observe these fields in other wavebands.

\subsection{\label{sec:Radio_AGN}Radio searches for AGN}


The first step in searching for (or excluding) AGN counterparts of \Fermi-LAT unassociated sources is to consult catalogs of radio sources.  Almost all radio AGN candidates of possible interest are detected either in the NRAO VLA Sky Survey \citep[NVSS;][]{nvss} or the Sydney University Molonglo Sky Survey \citep[SUMSS;][]{sumss}.  NVSS covers the entire $\delta > -$40$\arcdeg$ sky and provides interferometric flux density measurements at 1.4~GHz.  SUMSS covers the remainder of the sky and offers interferometric measurements at 0.843~GHz.  

In order to discover AGN counterparts, which are radio sources with compact, flat-spectrum cores, we follow the approach of the Cosmic Lens All Sky Survey \citep[CLASS;][]{class1,class2} and the Combined Radio All-Sky Targeted Eight GHz Survey \citep[CRATES;][]{crates}---both of which have been shown \citep{LBAS,1LAC} to include substantial numbers of radio counterparts of LAT blazars---and pursue 8.4~GHz follow-up interferometry of blazar candidates. The \Fermi-motivated VLA programs AH976 \citep{gaps} and AH996 (in progress) obtained such data for several hundred sources, and $\sim50$ of these appear as ``affiliations'' (i.e., candidate counterparts for which quantitative association probabilities could not be computed) in the 1LAC catalog. 108 new associations, including the ``figure of merit'' value for each association \citep{1LAC}, were determined by these VLA follow-up programs. 

Recently, serendipitous  radio identification surveys  of 1FGL sources have  been  independently  carried  out  using  the recently released Australia Telescope  20~GHz radio source catalog  \citep{Murphy10} which contains entries  for 5890 sources observed at $\delta<$ 0$\degr$.  Cross-correlation between the 1FGL  source  list and  the AT20G catalog has been performed by \citet{Mahony10} and  \citet{Ghirlanda10}.  In particular,  \citet{Mahony10} find correlated radio sources for  233 1FGL sources and, based on Monte Carlo tests, they infer that 95\% of these matches are genuine associations. While most of these radio detections are not identified or classified as specific object types, nine of the sources are considered likely to be AGNs (based on their properties at radio frequencies). \citet{Ghirlanda10} obtain a similar number of matches  with the AT20GHz catalog ($\sim$230) and propose eight of the same sources as likely AGN. All but one of these new associations were also found by the VLA observing program. 

To date, radio observations of sources in the 1FGL unassociated sample have produced 109 new AGN associations. These new AGN associations are included in Table~\ref{tbl:assoc_table}. In addition, the 1LAC catalog documented 57 other AGN associations with 1FGL unassociated sources. These have also been added to the table. Where possible, names have been adjusted to be consistent with the NASA/IPAC Extragalactic Database\footnote{http://ned.ipac.caltech.edu/} nomenclature.

\subsection{\label{sec:Radio_psr}Radio searches for pulsars}

Of the 56 $\gamma$-ray emitting pulsars identified in 1FGL, 32 were detected by folding the $\gamma$-ray data using timing solutions from observations of known radio pulsars. These ephemerides have been collected by a global consortium of radio astronomers who dedicate a portion of their time toward this effort \citep{DAS08}. The 32 pulsars (23 young and nine MSPs) had all been discovered in the radio band prior to their detection by the LAT. Since the release of the 1FGL catalog, twelve more of the 1FGL sources have been found to have $\gamma$-ray pulsations by using ephemerides of known radio pulsars. While eight of the twelve were associated in 1FGL with a pulsar or a SNR/PWN, four were listed as unassociated sources in the catalog (two young pulsars and two MSPs). These new associations are listed in Table~\ref{tbl:assoc_table}.

In addition to folding data using the properties of known radio pulsars, a promising technique for identifying unassociated sources is searching for previously unknown radio pulsars that might be powering the $\gamma$-ray emission.  This technique was used on many of the EGRET unidentified sources \citep[for example]{Champion2005,Crawford2006,Keith2008} with modest success, because the error boxes were many times larger than a typical radio telescope beam. With the LAT, the unassociated source localizations are a much better match to radio telescope beam widths and generally each can be searched in a single pointing. 

Thus far, over 450 unassociated LAT sources, mostly at high Galactic latitudes, have been searched by the \Fermi~Pulsar Search Consortium \citep[PSC;][]{Ransom2011} at 350, 820, or 1400 MHz. The target lists for these searches were selected from the LAT unassociated sources, with preference for those that displayed low variability and a spectrum consistent with an exponential cutoff in the few GeV range, as seen in the identified $\gamma$-ray pulsar population \citep{pulsarcat}. This program has resulted in the discovery of 32 previously unknown radio pulsars (one young pulsar and 31 MSPs) \citep{Ray2010,Keith2011,Cognard2011,Bangale2011} that are included in Table~\ref{tbl:assoc_table}. Of these 32 new pulsars, 14 also show pulsations in $\gamma$ rays. There is no obvious correlation between the $\gamma$-ray and radio fluxes of these pulsars, so additional discoveries can be expected as fainter unassociated 1FGL sources are searched. 

Searches by the PSC continue.  In the Galactic plane, high dispersion measures and sky temperatures demand higher frequency observations with concomitantly smaller beam sizes. Young, energetic pulsars can be very faint in the radio \citep{Camilo2002c,Camilo2002d}. Nevertheless, we expect that deep observations will continue to turn up more discoveries of radio pulsars in unassociated 1FGL sources in the near future. 

To summarize, radio observations of sources in the 1FGL unassociated sample have produced 36 new pulsar associations. 18 of these pulsars are considered firm identifications due to the detection of pulsations in the LAT data. 

\subsection{\label{sec:xrayobs}X-ray observations of unassociated source fields}

To look for additional possible counterparts we cross-correlated the list of unassociated 1FGL sources with existing X-ray source catalogs. We stress that the resulting compilation has no claim of completeness since the match with cataloged X-ray sources depends on the serendipitous sky coverage provided by the X-ray observations, and the integration time of the observation. While it is possible that  candidate X-ray counterparts to the LAT unassociated sources may be singled out on the basis of, e.g., their brightness and/or spectral properties, most will be recognized only through a coordinated multiwavelength identification approach (which is beyond the scope of this paper).  

To begin, we considered the 2XMM source catalog derived from pointed {\it XMM}-Newton observations \citep{XMMserendip}. The fourth incremental release of the catalog (2XMMi) includes 191,870 unique X-ray sources extracted from all {\it XMM}-Newton observations that were public as of 2008 May 1, i.e. performed through the end of 2007 April. We cross-correlated the LAT source lists with the 2XMMi catalog, using a  cross-correlation radius equal to the semi-major axis of the 95\% confidence ellipse of the LAT source, and found that 40 of the 1FGL unassociated fields contained 2XMMi detections. Of these 40, four had been found to be associated with AGN by the radio follow-up observations (Section~\ref{sec:Radio_AGN}). 

By looking at the {\it XMM}-Newton observation log (up to 2011 February 27) we can estimate the potential increase in the number of matches that could occur if we were to use the longer observational database. For the cross-correlation we used the LAT 95\% semi-major axis and compared against a radius equal to the sum in quadrature of the EPIC camera radius (15\arcmin) and the r95 positional uncertainty for the X-ray source. Here we found another 17 fields of unassociated LAT sources that have been observed by {\it XMM}-Newton, either serendipitously or pointed.

In addition, we cross-correlated with the {\it ROSAT} All Sky Survey Bright and Faint Source Catalogs \citep{Voges99,Voges00} and with the {\it ROSAT} catalog of pointed Position-Sensitive Proportional Counter (PSPC) \citep{Voges94} and High Resolution Imager (HRI) \citep{Voges2000} observations. These found 101 unassociated source fields with ROSAT counterparts, 15 of which were also found in the 2XMMi correlation. These results are summarized in Table~\ref{tbl:assoc_table}, and show that a preliminary X-ray screening provides potential X-ray counterparts for about 20\% of all the \Fermi-LAT unassociated sources. These possible X-ray counterparts are obviously prime targets for multiwavelength follow-up observations.  

We  also compared the 1FGL unassociated source list with recently published catalogs of hard X-ray/soft $\gamma$-ray sources. These are the {\it Palermo Swift-BAT Hard X-ray Catalog} \citep{SwiftBATcat} which is a compilation of 754 sources detected by the BAT instrument in its first 39 months of operations, and the 4th {\it IBIS/SGRI} Soft $\gamma$-ray Survey Catalog \citep{4thIBIScat} which includes a total of 723 sources. Both catalogs contain flux and spectral information and provide likely source identification/classification. We cross-correlated the {\it Swift}-BAT and {\it IBIS/SGRI} source catalogs using the nominal 1FGL position and 95\% semi-major axis as before, and found 11 new associations and 8 new X-ray detections (cases where the candidate X-ray counterpart is not a known $\gamma$-ray emitter). The 1FGL 95\% positional error is larger than the positional errors on sources in both catalogs. These results are also included in Table~\ref{tbl:assoc_table}, where we give the {\it Swift}-BAT and Integral-IBIS/SGRI source identifications and proposed catalog classifications. 

We note that a preliminary cross-correlation of the LAT unassociated sources with the ROSAT sources has been performed by \citet{Stephen2010}. However, they used only the ROSAT Bright Source Catalog as a reference and found, on statistical grounds, that 60 of the 77 correlated positions should be genuine associations. However, they provide likely associations for only 30 of the correlations, those with counterparts within $160\arcsec$. Table~\ref{tbl:assoc_table} lists 26 of these correlated sources. Three are not listed because the counterpart source type is not a known $\gamma$-ray emitter. The final source is not listed because no counterpart name was provided.

A survey of 21 fields of unassociated 1FGL sources was carried out by \citet{Mirabal2009_arXiv} using data from the Swift science archive. This investigation indicated X-ray detections for seven LAT unassociated sources based on positional correlation with Swift-XRT sources and the likelihood the source is a member of a $\gamma$-ray emitting class. Three of these are unique to this investigation and have been included as X-ray detections. In addition, \citet{Mirabal2010_arXiv} proposed nine possible associations for unassociated LAT sources at $|b| > 25\degr$ in the 3000 square degree ``overlap region'', a region covered by various radio surveys and by the Sloan Digital Sky Survey. Associations and detections from both these investigations are included in Table~\ref{tbl:assoc_table}.

The unassociated source 1FGL~J1958.9+3459 appears to be nearly coincident with the HMXB Cygnus X-1, which was recently reported as an {\it AGILE} source \citep{2010ATEL2715}. 
While this source meets the criteria to claim a positional association, there is no clear evidence that the source detected by the LAT is Cyg X-1. In addition, the source 1FGL J1045.2$-$5942 is positionally coincident with the luminous blue variable (LBV) star, Eta Carinae ($\eta$ Car). While X-ray observations of $\eta$ Car show a 5.54 year periodicity, the $\gamma$-ray flux remained constant during the most recent X-ray minimum in 2008 December -- 2009 January. However, due to its unusual $\gamma$-ray spectrum this 1FGL source is still believed to be associated with $\eta$ Car \citep{LATetaCar}.

To date, X-ray observations have led to positional associations with ten AGNs, seven HMXBs, one SNR and the LBV Eta Carinae. An additional 110 sources have X-ray detections that are excellent targets for follow-up multiwavelength observations. These associations can be found in Table~\ref{tbl:assoc_table}.

\subsection{\label{sec:TEVobs}TeV observations of unassociated sources}

\Fermi-LAT spectra have been shown to be good predictors of TeV emission, with 55 1FGL sources having very high energy (VHE) counterparts \citep{LATTeVAGN,LATPSR1907}. The energy range from $\sim$50~GeV to $\sim$300~GeV is the only range where the LAT data directly overlap with other instruments. 

The LAT team has provided recommendations for follow-up observations of a number of hard-spectrum sources -- including unassociated hard-spectrum sources -- that may have VHE counterparts. Coordinated follow-up observations in the TeV regime have been useful in identifying LAT-detected AGNs \citep[see e.g.][]{LATTeVAGN,2010ATEL2916}. In addition, LAT sources have been identified as SNRs by comparing the extension in the LAT data to the VHE emission by using the same procedure as was used for W51C \citep{LATW51C}, W44 \citep{LAT_W44} and IC443 \citep{LATIC443}. This search has yielded two more identified SNRs, W28 \citep{LAT_W28} and W49B \citep{LATW49B}. These associations can be found in Table~\ref{tbl:assoc_table}.

We cross-checked the 1FGL unassociated source list with the list of TeV sources from TeVCat\footnote{http://tevcat.uchicago.edu} and current publications. We consider a source to be coincident with a LAT source if its extension overlaps with the 95\% confidence ellipse of the LAT source. We find nine TeV sources that are coincident with 1FGL unassociated sources (Table~\ref{tbl:TeV_candidates}).  Note that the 1FGL association process did not assign an association to a coincident TeV source if that TeV source had no identification in another waveband. Pismis 22 and W43 are possible (but not confirmed) associations with the TeV source.

One source of note is 1FGL J1702.4$-$4147 which lies on the emission ``tail'' of the elongated VHE source, H.E.S.S. J1702$-$402. The nearby pulsar PSR J1702$-$4128 lies at the edge of the TeV $\gamma$-ray emission and would provide enough spin-down energy loss to produce the observed VHE emission via an extremely asymmetric PWN \citep{HESSsurvey08}. Hence the pulsar is considered to be a likely counterpart to the LAT source. To date there has not been a high significance detection of pulsations in the LAT data. An additional interesting source, 1FGL J1839.1$-$0543 is positionally coincident with HESS J1841$-$055, one of the most extended (1 deg in diameter) H.E.S.S. unidentified sources. Because of the high density of potential counterpart sources in this low-latitude region, there are multiple possible associations for the VHE source (2 SNRs, 3 high spin-down PSRs, 1 XRB) \citep{HESSsurvey08}. Given its high TeV $\gamma$-ray flux, it is considered a good candidate for LAT detection \citep{HESS_Fermi2009}.

\subsection{\label{sec:latobs}Association of LAT sources using only LAT data}

A small number of sources have been associated or identified since the release of the 1FGL catalog by using LAT data alone. Of the 56 pulsars listed in 1FGL, 24 were discovered using blind frequency searches \citep{BSP} for $\gamma$-ray pulsations from the bright unassociated sources. These are typically young pulsars, for which the solid angle of the radio beam is likely to be much smaller than the $\gamma$-ray one \citep{pulsarcat}. As a result of this geometry, many unassociated sources are likely to be young, radio-quiet pulsars that will never be found in radio searches. Since the release of the 1FGL catalog, another blind search pulsar, PSR~J0734$-$1559 has been discovered in an unassociated LAT source. This association has been included in Table~\ref{tbl:assoc_table}.



One new HMXB has also been detected in the LAT data \citep{2011ATEL3221}, though in 1FGL the source was associated with the SNR G284.3$-$01.8. This is the first of its type to be discovered in $\gamma$ rays. These new associations are also included in Table~\ref{tbl:assoc_table}.

\section{\label{sec:discuss}Discussion}

The follow-up multiwavelength associations efforts discussed in Section~\ref{sec:newassoc} have resulted in 177 new extragalactic source associations (all AGN), and 52 new Galactic associations (one source has both a Galactic and extragalactic association).  When we compare these new associations with the expected 1FGL source distributions discussed in Section~\ref{sec:lognlogs}, the estimated numbers of sources that have not yet been associated are reduced to 182 for extragalactic sources and 219 for Galactic sources. 

To test the two classification algorithms and to estimate the efficiency for identifying the different source classes, we compare the results to the new source associations in Section~\ref{sec:newassoc}, first looking at individual methods, and then the combined results. In addition, we consider how the results match the Log{\it N}-Log{\it S} analysis.

\subsection{\label{sec:Valid_separate}Validating the classification results from the separate methods}

We separately compared the results of the Classification Tree analysis and the Logistic Regression analysis to the new source associations in Section~\ref{sec:newassoc}. Altogether, the new source associations include 177 new AGN associations as well as 37 new pulsar associations that are divided into 2 categories: 20 objects for which pulsations have been detected in $\gamma$ rays (which we will refer to as ``new pulsar detections'') and 17 objects for which pulsations have been detected only in the radio (which we will refer to as ``new pulsar candidates''). We will not use new associations from other source types (HMXB, PWN, SNRs) for this validation.

For AGN, we find that 126 sources are correctly classified as AGN candidates by the CT analysis (efficiency: 71\%), 11 were classified as pulsar candidates (false negative: 6\%), while the remaining 40 sources were considered still unclassified (23\%). The same comparison for the LR analysis gave 142 sources correctly classified as AGN candidates (efficiency: 80\%), 7 sources classified incorrectly as pulsar candidates (false negative: 4\%), while the other 28 sources remained unclassified (16\% of the sample). 

For pulsars, we noticed a very different performance between new pulsar detections and new pulsar candidates. For the 20 sources detected as pulsars by the LAT, the CT analysis correctly classifies 14 pulsars (efficiency: 70\%), mis-classifies one source (5\%), and leaves the remaining sources unclassified (25\%). The LR analysis correctly classifies 11 pulsars (efficiency: 55\%), mis-classifies one source (5\%), and leaves the remaining sources unclassified (40\%). On the other hand, for the new radio pulsar candidates, the CT analysis correctly classifies only 3 objects as pulsars (efficiency: 18\%), mis-classifies 8 objects as AGN (false negative: 47\%) and leave the 6 remaining objects as still unassociated (35\% of the new pulsar candidates). The LR analysis correctly classifies 4 pulsar candidates (efficiency: 23.5\%), mis-classifies four sources (23.5\%), and leaves the remaining nine sources unclassified (53\%). 

These results are interesting, as the definition of the pulsar fiducial threshold in the LR analysis appeared that it might over-estimate the pulsar candidates. However, the Logistic Regression actually has a somewhat poorer success rate for finding new pulsars and pulsar candidates than the Classification Tree analysis. Looking more closely at the 1FGL properties of the misclassified pulsar candidates, we found that twelve of the 17 new pulsar candidates have only upper limits for the 300 MeV $-$ 1 GeV band. In contrast, 80\% of the new pulsar detections were significantly detected in this portion of the LAT spectrum. This difference in characteristics for the two pulsar groups may indicate the need for additional criteria when selecting sources for follow-up observations.




\subsection{\label{sec:Valid_combined}Validating the results from the combined classifications}

We can also compare the new associations to the combined classifications defined in Section~\ref{sec:CombClass}. Of the 214 newly-associated AGN and pulsars from Section~\ref{sec:newassoc}, 171 sources (151 new AGN, 16 new pulsar detections, and 4 new pulsar candidates) match the classification given by the combined analysis, and 26 sources (15 new AGN, 1 new pulsar detection, and 10 new pulsar candidates) are in direct conflict with the classification source type. This gives an efficiency of 85\% for AGN classification and 80\% for classification of new pulsar detections, but only 59\% for new pulsar candidates. 17 of the newly associated sources are unclassified by either method, and only one source has conflicting source classification. The one conflicting source turns out to be a new pulsar candidate that also has an AGN association, suggesting the LAT source could be the sum of these two objects. The overall efficiency for this combined sample is $\sim$80\%, comparable to the value we were seeking when we set the fiducial values for the two methods. The combined sample has a false negative rate of $\sim$12\%.

\begin{figure}[!ht]
\centering
\includegraphics[width=0.8\textwidth]{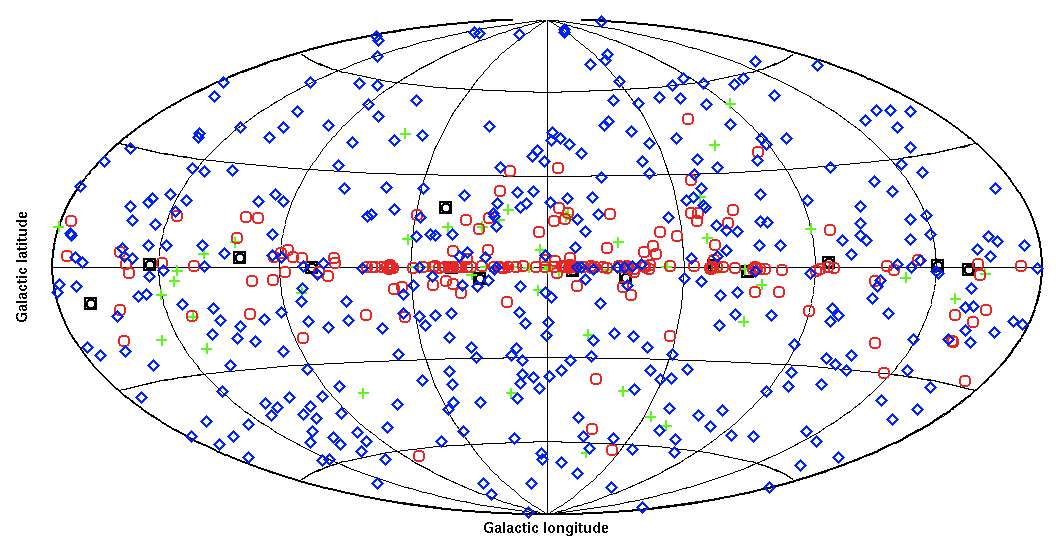}
\caption{The spatial distribution of the combined classification sample, in Galactic coordinates. Sources are classified as AGN candidates (blue diamond), pulsar candidates (red circles), unclassified (green crosses), or in conflict (black squares).
\label{fig:Class_galactic}}
\end{figure}

The spatial distributions of the newly classified sources give us the opportunity to cross check our results. Figure~\ref{fig:Class_galactic}  shows the spatial distribution of all the sources. Notice that both the AGN and pulsar distributions are as expected, even though we have not used the Galactic latitude as an input to either classification method. The pulsar candidates are mainly distributed along the Galactic plane, with a few high-latitude exceptions that suggest additional nearby MSPs, while the AGN candidates are nearly isotropically distributed on the sky. 

From this we can conclude that using only the $\gamma$-ray properties of the \Fermi~LAT sources, and the firm associations of the 1FGL, we were able to develop a prediction for AGN and pulsars classification that nearly matches our expectations (i.e. pulsar candidates are not variable, have a curved spectrum and are mainly distributed along the Galactic plane, while AGN candidates are mostly extragalactic, variable sources). In all, the efficiency of the combined classification methods at classifying new AGNs is high, with a low rate of false negatives, while the efficiency for new pulsar candidates is much lower than expected. 


AGN and pulsars are not the only $\gamma$-ray source classes known or expected, but the less-populous source types are hard to classify using the techniques described here because the training samples are too small in 1FGL.  With time, those training samples will likely grow, and we may be able to extend this analysis to additional classifications.


\subsection{\label{sec:Valid_logNlogS}Comparison to the Log{\it N}-Log{\it S} predictions}

In addition, we can check to see how the classification results compare to the predictions made by the Log{\it N}-Log{\it S} analysis in Section~\ref{sec:lognlogs}. For this comparison, we consider a pulsar classification to be indicative of a Galactic source, and an AGN classification to be indicative of an extragalactic source. 

Since 229 of the unassociated sources now have associations, we will consider only those 401 that remain unassociated. Thirty-three of these are unclassified by either technique, and 13 have conflicting classifications. The number sources for low latitudes and high latitudes for the remaining 355 are shown in Table~\ref{tbl:sourcedistribution}. 
At high latitudes, the observed numbers of both Galactic and extragalactic sources are consistent with the numbers expected from the Log{\it N}-Log{\it S} analysis.  In contrast, at low latitudes, the number of Galactic sources is about one-third lower than expected, and the number of extragalactic sources is  higher than expected.  It is clear that the group of sources that is hardest to associate is those of Galactic origin at low latitudes, likely due to the presence of a population of spurious sources in that region in the 1FGL catalog.

\begin{figure}[!ht]
\centering
\includegraphics[width=0.4\textwidth]{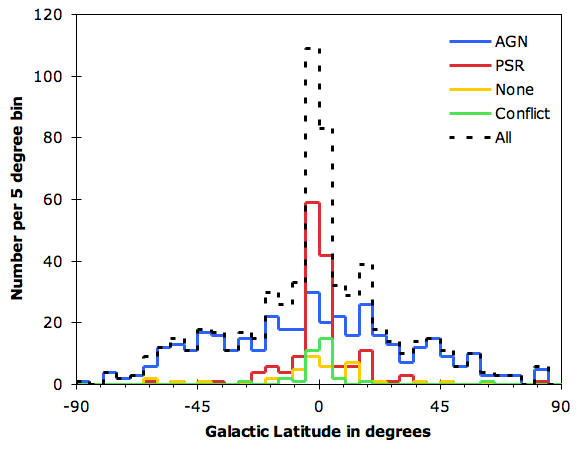}
\includegraphics[width=0.4\textwidth]{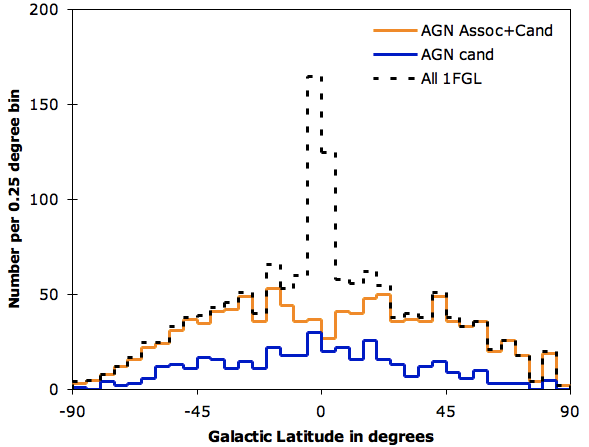}
\caption{Left:  Distribution of classified sources binned by Galactic latitude, with AGN in blue, pulsars in red, unclassified sources in green and sources with conflicting classification in yellow. The dashed line in the total distribution.  Right: Distribution of  AGN candidate binned by Galactic latitude. The orange line is the sum of the 1FGL AGN associations plus the sources classified as AGN candidates (blue line). The dashed line is the distribution for all 1FGL sources.
\label{fig:Class_histo}}
\end{figure}

Figure~\ref{fig:Class_histo} (left panel) shows the latitude distribution of the classified unassociated sources based on classification type. 
If we combine the AGN candidate population with the 1FGL sources that already have AGN associations (Figure~\ref{fig:Class_histo}, right panel), we find that the shape of the AGN distribution matches reasonably with that predicted by the model, though there is still an excess at low Galactic latitudes.

\subsection{Unassociated sources at low Galactic latitudes}

\begin{figure}[!ht]
\centering
\includegraphics[width=0.8\textwidth]{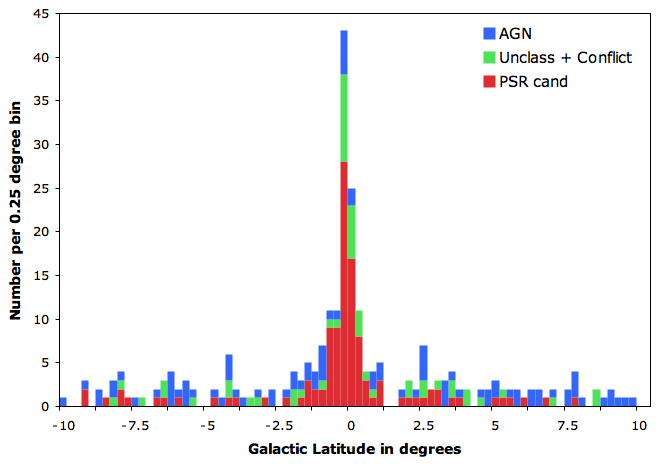}
\caption{The 1FGL unassociated sources in the central few degrees of the Galaxy can be mostly separated into those classified as pulsars (red) and those that have conflicting classifications or were unable to be classified (green). The few remaining sources were classified as AGN (blue).
\label{fig:Class_profiles}}
\end{figure}

It is the unassociated sources in the central 20$\degr$ of Galactic latitude ($|b|<10\degr$) that may hold clues to the content of the narrow Galactic ridge population at $|b| < 0\fdg5$, $|l| < 60\degr$ discussed in Section~\ref{sec:fluxdist}. To investigate these sources, we separate the pulsar candidates from the other types of sources (Figure~\ref{fig:Class_profiles}) and consider the distribution. Where the full set of unassociated sources appears to indicate an unreasonably narrow scale height of $\sim$50 pc for the population, the latitude distribution of pulsar candidates is somewhat closer to expectations, implying a scale height of $\sim$85 pc. This value is one-third the scale height of LAT-detected $\gamma$-ray pulsars \citep{pulsarcat}.

For $\gamma$-ray sources, this population scale height suggests instead that Population I objects such as SNRs, with a scale height of $\sim$100 pc \citep{Lamb1997}, are likely significant contributors to the 1FGL sources. In the 1FGL catalog, 44 sources were associated with or identified as SNRs, and this paper associates six more (Table~\ref{tbl:assoc_table}), giving a total of 50 SNR associations. Considering that only 63 1FGL sources were associated/identified with pulsars, it is clear that both source types are significant contributors. The classification method used here does not consistently label SNRs as pulsar candidates. Of the six new SNR associations, five are classified as pulsar candidates, and the sixth is classified as an AGN. In the future it may be useful to consider the SNR source class separately as an input to such classification analyses. 

This central portion of the Galaxy is also the region that has most of the sources that are either unclassified or have conflicting classifications. Of the 257 unassociated sources in this region, 22 have no classification, and 29 have conflicting classifications between the two methods. Of these 51 unclassified or conflicting-classification sources, eight have new associations. The source types are varied; two MSPs, one young pulsar, three HMXBs and two AGN. It is clear that not all of these sources can be spurious, however it is unlikely the remaining 43 are all real detections.


\subsection{Informing future follow-up observations}

The results of the classification analyses demonstrate that source properties measured with the \Fermi-LAT can provide important guidance on what types of follow-up observations are likely to be fruitful for many of these unassociated sources. The emphasis in follow-up observations of LAT sources has been on radio imaging and timing observations for a large number of sources, as well as targeted X-ray observations for sources of interest (e.g. flaring sources or new radio pulsar candidates). In addition, there is an on-going program to observe all the bright, well-localized \Fermi-LAT unassociated sources with {\it Swift}\footnote{http://www.swift.psu.edu/unassociated/} that may add important new insights into these sources as a group.

In Table~\ref{tbl:gamma_table}, the last 4 columns show what follow-up observations are recommended in several wavebands. Obviously, sources classified as likely blazars would benefit from radio searches for flat-spectrum sources within the LAT error ellipse. Low-frequency radio timing is recommended for the likely pulsars. X-ray observations of likely pulsars can give timing observers seed locations at which to search for pulsations in both radio and LAT data \citep{Caraveo2009}.  Still-unassociated sources that may benefit from such observations have been flagged in the 1FGL unassociated source list with the appropriate observation type. These are suggestions; it is highly likely that some of the sources are misclassified. Also, a number of the follow-up observations discussed here have yielded no new associations for some of the observed unassociated sources. 

We strongly recommend additional joint analyses of LAT and ground-based VHE $\gamma$-ray data for very low-latitude ($|b| < 0\fdg5$) \Fermi~sources. Together, these may give insight into whether or not a population of SNRs can account for a significant number of the 1FGL unassociated sources along the Galactic ridge. Sources for which this type of analysis is recommended are indicated in the 1FGL unassociated source list (Table~\ref{tbl:gamma_table}).

\subsection{\label{sec:bslunids}Remaining BSL unassociated sources}

Follow-up observations like those discussed in the previous section have made a significant impact, increasing the associated fraction for the 1FGL catalog from $\sim$56.5\% to $\sim$71\% in only a little more than one year. But we can look farther back to the \Fermi-LAT Bright Source List \citep[BSL][]{BSL}, the list of 205 high significance ($>10\sigma$) LAT sources detected in the first three months of the \Fermi~mission.  Of the 37 sources listed as unassociated in the BSL, ten now have pulsar identifications or associations, while eight have new AGN associations. In addition, seven have been associated with other $\gamma$-ray source types such as SNRs or HMXBs.  These associations bring the BSL association rate up to 94\%. 

Of the remaining twelve BSL unassociated sources, five lie in the Galactic ridge and should be considered with caution. Here we look more closely at these 12 sources:

\begin{itemize}

\item {\bf 1FGL\,J0910.4$-$5055} (0FGL\,J0910.2$-$5044) -- This Galactic plane transient ($l,b = 271\fdg7,-1\fdg96$) flared once in October 2008, an event lasting 1--2 days with peak flux ($E>100$ MeV) of 1$\times$10$^{-6}$ \pflux \citep{2008ATEL1788}. (1FGL reported a peak flux for this source of 1.97$\times$10$^{-7}$ \pflux, but that was averaged over one month.) Recent figure-of-merit analysis by \citet{Murphy10} has given this source a $>80$\% probability of being associated with the radio source AT20G\,J091058-504807. Our analysis does not find such an association, though \citet{Mirabal2009_arXiv} does show an association with the likely blazar {\it Swift} J0910.9$-$5048.

\item {\bf 1FGL\,J1311.7$-$3429} (0FGL\,J1311.9$-$3419) -- Besides being associated with 3EG\,J1314$-$3431, this very bright high-Galactic latitude  ($l,b = 307\fdg7,28\fdg2$) source is not variable and has a spectrum with a high-energy cutoff very similar to a pulsar. To date, searches for both $\gamma$-ray and radio pulsations from this source have been unsuccessful \citep{Ransom2011}. 

\item {\bf 1FGL\,J1536.5$-$4949} (0FGL\,J1536.7$-$4947) -- With no significant variability, this persistently bright mid-latitude  ($l,b = 328\fdg2,4\fdg8$) source is the only unassociated BSL source to have conflicting classifications. This source has a moderately curved spectrum with a high-energy cutoff.

\item {\bf 1FGL\,J1620.8$-$4928c} (0FGL\,J1622.4$-$4945) -- This moderately bright source in the Galactic ridge  ($l,b = 333\fdg9,0\fdg4$) is spatially coincident with the {\it AGILE} detection 1AGL J1624-4946, and has a spectrum with a sharp spectral break at $\sim$ 3 GeV. There is no evidence at this time for pulsations from this source in either $\gamma$ rays or radio \citep{Ransom2011}.

\item {\bf 1FGL\,J1653.6$-$0158} (0FGL\,J1653.4$-$0200) -- Another non-varying source with a pulsar-like spectrum, this high-latitude ($l,b = 16\fdg6,24\fdg9$) source is associated with 3EG\,J1652$-$0223. Both classification methods call this source a pulsar candidate.

\item {\bf 1FGL\,J1740.3$-$3053c} (0FGL\,J1741.4$-$3046) -- The position for this non-varying c-source in the Galactic ridge ($l,b = 357\fdg7,-0\fdg1$) moved far enough between the two publications that the two detections are not formally associated. However, we recall that for pulsars in the plane, the 1FGL error ellipse appears to underreport the systematic error. As the BSL source lies just outside the 1FGL 95\% confidence contour, we consider the two detections to be related. 

\item {\bf 1FGL\,J1839.1$-$0543c} (0FGL\,J1839.0$-$0549) -- This bright source in the Galactic ridge  ($l,b = 26\fdg4,0\fdg1$) has a highly curved spectrum and does not vary. Both classification methods call this source a pulsar candidate.

\item {\bf 1FGL\,J1842.9$-$0359c} (0FGL\,J1844.1$-$0335) -- This Galactic ridge source  ($l,b = 28\fdg4,0\fdg1$) had a source significance of 10$\sigma$ in the first three months, but after eleven months that value has increased only slightly, to 10.9$\sigma$ (unless variable, a source should have twice the significance after nearly four times the livetime). Since 1FGL found this source to be non-varying, it is unlikely that the low flux in 1FGL is due to variability effects. Instead, it appears that the longer data set has separated the BSL source into multiple components, leaving the coincident 1FGL source at a lower-than-expected significance.

\item {\bf 1FGL\,J1848.1$-$0145c} (0FGL\,J1848.6$-$0138) -- Another source in the Galactic ridge  ($l,b = 31\fdg0,-0\fdg1$), its spectrum is consistent with a power-law, and the source may be related to a TeV source (see Table~\ref{tbl:TeV_candidates}).

\item {\bf 1FGL\,J2027.6+3335} (0FGL\,J2027.5+3334) -- A bright, mildly variable source that lies near the Galactic plane ($l,b = 73\fdg3,-2\fdg9$), this source is associated with the EGRET source 3EG\,J2027+3429. In the 1FGL catalog, this source flux peaked at $3.4 \times10^{-7}$ \pflux~ with a significance $> 10\sigma$ in a single month. The spectrum has a large curvature index, indicating that it is not a simple power-law. Although the variability seems to indicate a possible AGN, both classification methods consider it a likely pulsar.

\item {\bf 1FGL\,J2111.3+4607} (0FGL\,J2110.8+4608) -- While this source near the Galactic plane  ($l,b = 88\fdg3,-1\fdg4$) is highly significant, it is flagged in the 1FGL catalog as having the flux measurement that is sensitive to changes in the diffuse model. Even so, the spectrum is moderately pulsar-like with a high-energy cutoff and there is no hint of variability. Both classification techniques consider this source a pulsar candidate.

\item {\bf 1FGL\,J2339.7$-$0531} (0FGL\,J2339.8$-$0530) -- This very hard source (spectral index = 1.99) lies at high Galactic latitude ($l,b = 81\fdg4,-62\fdg5$). While its 1FGL five-band spectrum suggests a blazar, neither classification method was able to classify this source.

\end{itemize}

While five of these are c-sources, only two (1FGL\,J1842.9$-$0359c and 1FGL\,J1848.1$-$0145c) appear to be questionable, though the possibility of a TeV component for the latter should be investigated. The two variable sources seem likely to be AGN. The $\gamma$-ray characteristics of the majority of the other unassociated BSL sources imply that these sources are bright, steady, and have curved spectra with high-energy cutoffs. With the exception of 1FGL\,J0910.4$-$5055, all these sources were included in the searches for radio pulsations, the same searches which have resulted in the discovery of ten new MSPs in unassociated BSL sources. Blind searches for pulsations in the $\gamma$-ray data have also been performed on these sources, with no success.

One source of interest is 1FGL 2017.3+0603, a BSL source that now has detected pulsations in the LAT data \citep{Cognard2011} as well as an AGN association with a high figure of merit (0.923) in 1LAC \citep{1LAC}. This source will require additional analysis to determine if the LAT flux is due solely to the pulsar, or is a combination of both counterparts.

\subsection{\label{sec:EGRETcompare}Comparing with EGRET unassociated sources}

Although the present paper is focused on the \Fermi-LAT unassociated sources, some insight about these sources may be found from the all-sky survey with EGRET on the {\it Compton Gamma Ray Observatory}.  Using the LAT results, with  higher sensitivity and better source locations, as a reference, we re-examine the sources of two EGRET catalogs: the 3EG catalog \citep{3rdCat} and the EGR catalog \citep{EGR}.  The two catalogs were based largely on the same data, while they used different models of the diffuse $\gamma$ radiation that forms a background for all sources.  In addition, noting those EGRET unassociated sources that remain unassociated in the 1FGL catalog offers the opportunity to recognize sources that remain interesting on a time scale of decades. 

\subsubsection{Experience with the EGRET Catalogs}

A comparison of the two EGRET catalogs with each other and with the LAT 1FGL catalog yields several observations and tentative conclusions, illustrated by specific examples.  Because more detailed information is available about the 3EG sources, many of these results emphasize that catalog, although similar considerations likely apply to the EGR analysis. 

\begin{enumerate}

\item The statistical error contours produced for EGRET underestimated the full uncertainty in the source localization.  The Crab, Vela, Geminga, and PSR~J1709$-$4429 pulsars, which are positively identified by $\gamma$-ray pulsations in both EGRET and \Fermi-LAT data, have positions outside the formal error contours, even at the 99\% level, as noted by \citet{3rdCat}.  For 3EG, but not for EGR, the CTA 1 and LSI +61$^\circ$ 303 sources, now firmly identified by LAT, also lie outside the 95\% error contours.  For this reason, we expand the list of plausible associations to include all those LAT sources whose position falls within the 99\% error contours.   As noted by \citet{3rdCat}, however, the EGRET source localizations were better at higher Galactic latitudes.  Even bright AGN such as 3C279 were typically found within the error contours.

\item The circular fit to the EGRET 95\% error contour was a poor approximation in many cases.  In addition to the 107 matches between 1FGL and 3EG reported based on the automated comparison using this circular fit, there are 21 more 1FGL sources found within the 95\% error contours of the detailed 3EG uncertainty contour maps \citep{3rdCat}.  Further, 19 1FGL sources are located between the 3EG 95\% and 99\% contours, making a total of 149 candidate associations, or 153 including the four bright pulsars identified by timing.  The 14 other 1FGL sources that lie just outside the 3EG 99\% contours are not included in this analysis, although they remain potential association. 

\item The adopted diffuse background model is important both close to and far from the Galactic plane.  For example, EGR J0028+0457 is confirmed by the LAT as the millisecond pulsar  PSR J0030+0451 \citep{PSR0030LAT} at a Galactic latitude of $-$57$^\circ$.  There was no corresponding source in the 3EG catalog, though a sub-threshold excess was seen \citep{DJT_pc}.  Conversely, 3EG J1027$-$5817, at a Galactic latitude of $-$1$^\circ$ is confirmed by the LAT as PSR J1028$-$5819 \citep{PSR1028LAT}, while the nearest EGR source is nearly 1$^\circ$ away with a 95\% error uncertainty of 0.22$^\circ$.  

\item Variability is an important consideration even for sources not associated with blazars.  The EGRET upper limit for radio galaxy NGC 1275, derived from Figure~3 of \citet{3rdCat}, lies nearly an order of magnitude below the LAT detection level \citep{LATNGC1275}.  3EG J0516+2320 was a bright solar flare from 1991 June \citep{Kanbach1993}.  The 1FGL catalog contains no solar flares. 3EG J1837$-$0423 \citep{Tavani1997} was one of the brightest sources in the $\gamma$-ray sky in 1995 June, at a flux (E$>$ 100 MeV) of (3.1 $\pm$ 0.6) x 10$^{-6}$ \pflux.  No 1FGL source is seen consistent in position with this source, even though the LAT would have detected a source nearly 2 orders of magnitude fainter \citep[see Figure 19 of][]{1FGL}. 

\end{enumerate}

Some of these lessons from the EGRET era have already been applied to the 1FGL catalog construction but are worth reiterating in any discussion of unassociated sources. 

\begin{enumerate}
\item  Unlike EGRET, the error contours for 1FGL include a systematic component of 10\% added to the statistical uncertainties.  This component was derived, however, from high-latitude AGN comparisons.  The EGRET experience suggests that the situation will be more difficult at lower Galactic latitudes.  The low-latitude LAT source associated with LS5039 based on periodicity, for example, has a measured position that lies at the 95\% uncertainty contour even after adding a 20\% systematic component \citep{LATLS5039}. 
This is an additional indication that the LAT positional uncertainties for sources in the Galactic plane are affected by the systematics discussed earlier. 
\item  Even without any additional systematic uncertainty, it should be remembered that at least 5\%, or more than 70, of the 1FGL sources probably have true counterparts that fall outside the 95\% contours. 
\item  Just as the EGRET catalog used ``C'' for potentially confused regions and ``em'' for possibly extended or multiple sources, the 1FGL catalog identifies sources with a ``c'' in the name or a numerical flag to indicate possible uncertainties due to the analysis procedure or the diffuse model. The LAT has confirmed $\sim$67\% of EGRET sources that do not carry either the ``C'' or ``em'' flags, but has only confirmed $\sim$38\% of the flagged sources. This experience with EGRET certainly suggests that such flags should be taken seriously. 
\end{enumerate}

\subsubsection {1FGL Unassociated Sources Remaining from the EGRET Era}

Despite the gap of more than a decade between the EGRET and LAT observations, a sizable number of the unassociated sources seen by EGRET are associated with 1FGL sources that are unassociated with known source classes.  Whether they are persistent or recurrent, such sources offer a high potential for multiwavelength studies. Table~\ref{tbl:assoc_table} includes 43 positional coincidences we have found between unassociated sources in the EGRET and 1FGL catalogs.  As noted above, the predominance of 3EG sources results at least partly from the additional information available compared to the EGR catalog, which did not include confidence contour maps. 


These unassociated sources are distributed widely across the sky.  Only 10 of the 43 have ``c'' designations in 1FGL, and all of these lie close to the Galactic plane toward the inner Galaxy, where the Galactic diffuse emission is brightest and any deficiencies in the model of the diffuse emission would have the greatest effect on properties of the 1FGL sources.  Although the EGRET localization uncertainties are large, the density of 1FGL sources away from the Galactic Plane is not so large that accidental coincidences are a significant problem. EGR J1642+3940 (1FGL J1642.5+3947) appears to be a special case.  It appeared in the EGRET data only after the end of the 3EG data set, but it has been seen by the LAT.  \citet{EGR} suggest an association with blazar 3C~345, although it might also be associated with Mkn 501 \citep{Kataoka1999}.  Although it is shown as unassociated in the 1FGL catalog, recent LAT analysis also suggests one or more blazars contribute to this source \citep{Schinzel2010}.

\section{\label{sec:conclusion}Conclusions}

As the \Fermi~mission matures, it is important to take a look at the successes of the early mission to help inform and improve the association and follow-up of new sources. The continued multiwavelength observations and ongoing statistical association efforts for the 1FGL unassociated sources have led to associations for 70\% of the entire catalog. Since the release of the catalog, 45\% of all the extragalactic sources expected from the Log{\it N}-Log{\it S} analysis have been associated. In addition, 47\% of the expected high-latitude Galactic sources have been associated. Together, this gives associations for nearly 82\% of all expected extragalactic and nearly 62\% of expected high-latitude Galactic sources. However, there are associations for only $\sim$ 38\% of the expected sources of Galactic origin that lie at $|b|<10\degr$. 

The significant improvement in sensitivity of the \Fermi-LAT relative to EGRET, the substantially improved positional errors, and the sky-survey viewing plan made possible by the large field of view of the LAT have generated an unprecedented data set and allowed the production of the deepest-ever catalog of the GeV sky. From that foundation, the astronomical community has worked in concert to discover new additions to known $\gamma$-ray source classes, as well as adding SNRs, PWNe, starburst galaxies, radio galaxies, HMXBs, globular clusters and a treasure trove of millisecond pulsars to the list. 

With that success as a backdrop, there are clearly lessons we have learned from the 1FGL catalog process and follow-up analyses, as well as from experience with previous missions:

\begin{enumerate}

\item As discussed in \citet{1FGL}, it is clear that there is room for improvement in the plane of the Galaxy, and especially the ridge. This region contributes numerous questionable sources to the catalog. However, if we set that region aside, essentially half of the remaining 1FGL unassociated sources in other regions now have associations, giving an overall association rate of $\sim$70\% for 1FGL. In contrast, the 3rd EGRET catalog had an association rate of only 38\%.

\item Follow-up campaigns to associate or identify the LAT-detected sources have been extremely successful. A total of 178 new blazar candidate associations have been made, primarily by looking for radio candidates within the LAT error ellipses and then re-observing at additional frequencies to determine if the source has the characteristic flat spectrum in the radio. Thirty-one new Galactic field MSPs have been discovered based on locations provided by the LAT for radio pulsations searches. 

\item By using the distribution of detected sources, we can model the 1FGL $\gamma$-ray sky and predict how many of each general source type we expect in the catalog. After taking into account the associations made since the release of the 1FGL catalog, this analysis indicates that, as the mission continues, we might expect to find associations for at least $\sim$200 more AGN (mostly at high Galactic latitudes) and $\sim$50 new pulsars (equally divided at high and low latitudes) among the unassociated 1FGL sources.

\item We have applied two analysis techniques to infer the likely classification for unassociated sources, based solely  on their $\gamma$-ray properties. The $\gamma$-ray properties of sources, while not being sufficient on their own to determine source type, can provide important information regarding the parent source classes. Using the information provided by the LAT to inform the selection of $\gamma$-ray sources and wavelengths for follow-up studies can reduce the labor intensive nature of such observations and increase the likelihood of finding a viable association candidate. A preliminary assessment of this process shows a success rate of $\sim$80\%.

\end{enumerate}


\section{Acknowledgments}
\acknowledgments The \Fermi~LAT Collaboration acknowledges generous ongoing 
support from a number of agencies and institutes that have supported both the development 
and the operation of the LAT as well as scientific data analysis. These include the National 
Aeronautics and Space Administration and the Department of Energy in the United States, 
the Commissariat \`a l'Energie Atomique and the Centre National de la Recherche 
Scientifique / Institut National de Physique Nucl\'eaire et de Physique des Particules in 
France, the Agenzia Spaziale Italiana and the Istituto Nazionale di Fisica Nucleare in Italy, 
the Ministry of Education, Culture, Sports, Science and Technology (MEXT), High Energy 
Accelerator Research Organization (KEK) and Japan Aerospace Exploration Agency 
(JAXA) in Japan, and the K.~A.~Wallenberg Foundation, the Swedish Research Council 
and the Swedish National Space Board in Sweden.

Additional support for science analysis during the operations phase is gratefully 
acknowledged from the Istituto Nazionale di Astrofisica in Italy and the Centre 
National d'\'Etudes Spatiales in France.

{\it Facilities:} \facility{\Fermi-LAT}.

\bibliography{Unassociated_sources}

\clearpage


\begin{deluxetable}{lccc}
\tabletypesize{\scriptsize}
\tablecaption{Spatial distribution of various source associations from the 1FGL and 1LAC catalogs
\label{tbl:1FGLdistro}}
\tablewidth{0pt}

\tablehead{
\colhead{Source}&
\colhead{Sources at}&
\colhead{Sources at}&
\colhead{Ridge\tablenotemark{a}}\\
\colhead{class}&
\colhead{$|b|>10\degr$}&
\colhead{$|b|<10\degr$}&
\colhead{sources}
}

\startdata
\textbf{Associated} & \textbf{670} & \textbf{151}  & \textbf{31} \\
\phm{3}AGN & 642 & 51 & 1 \\ 
\phm{3}Pulsars & 16 & 47 & 11 \\ 
\phm{3}SNRs/PWNe & 1 & 45 & 19 \\ 
\phm{3}Other & 11 & 8 & 0 \\
\hline
\textbf{Unassociated} & \textbf{373} & \textbf{257} & \textbf{88} \\
\phm{3}Non-c sources & 354 & 139 & 0 \\
\phm{3}c-sources & 19 & 118 & 88 \\
\enddata

\tablenotetext{a}{Here, the Galactic ridge is defined as sources with $|$b$|<$1$\degr$ and $|$l$|<$60$\degr$. This value is a subset of the previous column of $|$b$|<$10$\degr$ sources.}

\end{deluxetable}

\begin{deluxetable}{lccc}
\tabletypesize{\scriptsize}
\tablecaption{Expected vs. Observed source distribution
\label{tbl:sourcedistribution}}
\tablewidth{0pt}
\tablehead{

\colhead{Source}&
\colhead{Sources at}&
\colhead{Sources at}&
\colhead{Totals}\\
\colhead{types}&
\colhead{$|b|>10\degr$}&
\colhead{$|b|<10\degr$}&
\colhead{}
}

\startdata
Total detected & 1043 (71.9\%) & 408 (28.1\%) & 1451 \\
\phm{3}Associated & 670 & 151 & 821 \\
\phm{3}Unassociated & 373 & 257 & 630 \\
\hline
Extragalactic \\
\phm{3}Total from Log{\it N}-Log{\it S} & 972 & 88 & 1060 (73.1\%) \\
\phm{3}Associated & 650 & 51 & 701 \\
\phm{3}\textbf{Not Associated in 1FGL} & \textbf{322} & \textbf{37} & \textbf{359} \\
\phm{3}New Associations & 150 & 27 & 177 \\
\phm{3}New Classifications & 154 & 67 & 221 \\
\phm{3}\textbf{Log{\it N}-Log{\it S} Comparison} & \textbf{$-$18} & \textbf{+57} & \textbf{+39} \\
\hline
Galactic \\
\phm{3}Total from Log{\it N}-Log{\it S} & 71 & 320 & 391 (26.9\%) \\
\phm{3}Associated & 20 & 100 & 120 \\ 
\phm{3}\textbf{Not Associated in 1FGL} & \textbf{51} & \textbf{220} & \textbf{271} \\
\phm{3}New Associations & 27 & 25 & 52 \\
\phm{3}New Classifications & 31 & 103 & 134 \\
\phm{3}\textbf{Log{\it N}-Log{\it S} Comparison} & \textbf{+7} & \textbf{$-$92} & \textbf{$-$85} \\
\hline
\enddata

\tablecomments{Results from Log{\it N}-Log{\it S}  analysis, applied to the 1FGL source list.}

\end{deluxetable}

\clearpage

\begin{deluxetable}{lr}
\tabletypesize{\scriptsize}
\tablecaption{Ranking of the training variables for the Classification Tree.
\label{CT_ranking}}
\tablewidth{0pt}
\tablehead{
\colhead{Variable}&
\colhead{Importance}}

\startdata
Fractional Variability & 21.9\%\\
Hardness$_{45}$ & 16.0\%\\
Hardness$_{23}$ & 15.8\%\\
Spectral Index	& 13.0\% \\
Hardness$_{12}$ & 12.7\%\\
Hardness$_{34}$ & 11.8\%\\
Curvature	 & 8.8\%\\
\enddata

\tablecomments{~List of the training variables for the Classification
Tree: each variable is ranked according to its relevance in the discrimination
process, as computed by the CT algorithm.}

\end{deluxetable}

\clearpage

\begin{landscape}
\begin{deluxetable}{lrrrrccrcrcccccc}
\setlength{\tabcolsep}{0.05in}
\tabletypesize{\scriptsize}
\tablecaption{Summary of $\gamma$-ray properties and classification results
\label{tbl:gamma_table}}
\tablewidth{0pt}
\tablehead{
\colhead{1FGL Name}&
\colhead{R.A.}&
\colhead{Dec.}&
\colhead{$l$}&
\colhead{$b$}&
\colhead{Var\tablenotemark{a}}&
\colhead{BSL\tablenotemark{b}}&
\multicolumn{2}{c}{Class Tree}&
\multicolumn{2}{c}{Logistic Reg.}&
\colhead{Combined}&
\colhead{Radio}&
\colhead{Radio}&
\colhead{X$-$Ray}&
\colhead{TeV}\\
\colhead{}&
\colhead{(J2000)}&
\colhead{(J2000)}&
\colhead{}&
\colhead{}&
\colhead{}&
\colhead{}&
\colhead{Predict}&
\colhead{Class}&
\colhead{Predict}&
\colhead{Class}&
\colhead{Class}&
\colhead{Imaging}&
\colhead{Timing}&
\colhead{}&
\colhead{}}

\startdata
  J0000.8$+$6600c  & 0.209 & 66.002 & 117.812 & 3.635 &    &    & 0.64 &    & 0.08 &    &    &    &    &    &    \\
  J0001.9$-$4158  & 0.483 & $-$41.982 & 334.023 & $-$72.029 &    &    & 0.84 &  AGN  & 0.00 &  AGN  &  AGN  &  T  &    &    &    \\
  J0003.1$+$6227  & 0.798 & 62.459 & 117.388 & 0.108 &    &    & 0.77 &  AGN  & 0.06 &  AGN  &  AGN  &  T  &    &    &    \\
  J0004.3$+$2207  & 1.081 & 22.123 & 108.757 & $-$39.448 &    &    & 0.87 &  AGN  & 0.00 &  AGN  &  AGN  &  T  &    &    &    \\
  J0005.1$+$6829  & 1.283 & 68.488 & 118.689 & 5.999 &    &    & 0.78 &  AGN  & 0.00 &  AGN  &  AGN  &    &    &    &    \\
  J0006.9$+$4652  & 1.746 & 46.882 & 115.082 & $-$15.311 &    &    & 0.87 &  AGN  & 0.00 &  AGN  &  AGN  &  T  &    &    &    \\
  J0008.3$+$1452  & 2.084 & 14.882 & 107.655 & $-$46.708 &    &    & 0.77 &  AGN  & 0.00 &  AGN  &  AGN  &    &    &    &    \\
  J0009.1$+$5031  & 2.289 & 50.520 & 116.089 & $-$11.789 &    &    & 0.85 &  AGN  & 0.00 &  AGN  &  AGN  &    &    &    &    \\
  J0016.6$+$1706  & 4.154 & 17.108 & 111.135 & $-$44.964 &    &    & 0.87 &  AGN  & 0.00 &  AGN  &  AGN  &    &    &    &    \\
  J0017.7$-$0019  & 4.429 & $-$0.326 & 104.735 & $-$62.001 &    &    & 0.84 &  AGN  & 0.00 &  AGN  &  AGN  &    &    &    &    \\
\enddata

\tablecomments{Summary of the $\gamma$-ray properties of the 1FGL unassociated sources. The table includes flags for variability, predictor values for both Classification Tree and Logistic Regression analyses, combined classification, and flags indicating what type of follow-up observations are recommended. {\bf This table is published in its entirety in the electronic edition of the Astrophysical Journal. A portion is shown here for guidance regarding its form and content.}}

\tablenotetext{a}{T indicates the source was found to be variable in the 1FGL catalog analysis \citep{1FGL}.}
\tablenotetext{b}{T indicates the source was reported in the \Fermi-LAT Bright Source List \citep{BSL}.}

\end{deluxetable}
\end{landscape}

\clearpage

\begin{deluxetable}{lccr}
\tabletypesize{\scriptsize}
\tablecaption{List of the predictor variables for the LR model.
\label{tbl:LRpredicts}}
\tablewidth{0pt}
\tablehead{
\colhead{Variable}&
\colhead{Coefficient}&
\colhead{Standard Error}&
\colhead{p-value}}

\startdata
Intercept & -22.17 & 4.97 & $<$0.001\\
Fractional Variability & 10.61 & 1.49 & $<$0.001\\
Spectral Index & 11.30 & 2.47 & $<$0.001\\
Hardness$_{23}$ & -3.84 & 1.27 & 0.002\\
Hardness$_{34}$ & 8.14 & 1.53 & $<$0.001\\
Hardness$_{45}$ & 3.72 & 0.76 & $<$0.001\\ \hline
Hardness$_{12}$ & \nodata & \nodata & 0.242\\
glat & \nodata & \nodata & 0.333\\
glon & \nodata & \nodata & 0.144\\
\enddata

\tablecomments{Variables selected for the Logistic Regression analysis are listed at top. Those rejected are listed below the line.}

\end{deluxetable}


\begin{deluxetable}{lccc}
\tabletypesize{\scriptsize}
\tablecaption{Comparison of results for the classification techniques
\label{tbl:Class_summary}}
\tablewidth{0pt}
\tablehead{

\colhead{}&
\multicolumn{3}{c}{Classification}\\
\colhead{}&
\colhead{AGN}&
\colhead{Pulsar}&
\colhead{Unclassified}
}

\startdata
CT Totals & 304 & 160 & 166 \\
\phm{3}For $|b|>10\degr$ & 244 & 33 & 96 \\
\phm{3}For $|b|<10\degr$ & 60 & 127 & 70 \\
\hline
LR Totals & 368 & 122 & 140 \\
\phm{3}For $|b|>10\degr$ & 276 & 39 & 58 \\
\phm{3}For $|b|<10\degr$ & 92 & 83 & 82 \\
\hline
Combined & 386 & 177 & 53 \\
\phm{3}For $|b|>10\degr$ & 300 & 50 & 22 \\
\phm{3}For $|b|<10\degr$ & 86 & 127 & 31 \\
\hline
 & & LR Class & \\
CT Class & AGN & Pulsar & Unclassified \\
\phm{3}AGN &  \textbf{269} & \textit{2} & 33 \\
\phm{3}Pulsar & \textit{32} &  \textbf{72} & 56 \\
\phm{3}Unclassified & 94 & 31 &  \textbf{41} \\
\enddata

\tablecomments{Summaries for both high and low-Galactic latitude classification results for the Logistic Regression (LR) and Classification Tree (CT) techniques, as well as for the combined sample of classified sources (14 sources with conflicting classification are not included). In addition an inter-comparison of the two classification techniques is provided. Italics indicate conflicting results, while bold indicates agreement. All 630 1FGL sources are represented here.}

\end{deluxetable}

\clearpage

\begin{landscape}
\begin{deluxetable}{lrrrrcccclcc}
\setlength{\tabcolsep}{0.03in}
\tabletypesize{\scriptsize}
\tablecaption{Summary of follow-up observations and association efforts
\label{tbl:assoc_table}}
\tablewidth{0pt}
\tablehead{
\colhead{1FGL Name}&
\colhead{R.A.}&
\colhead{Dec.}&
\colhead{$l$}&
\colhead{$b$}&
\colhead{EGRET\tablenotemark{a}}&
\multicolumn{2}{c}{Association}&
\colhead{New\tablenotemark{b}}&
\colhead{Figure of\tablenotemark{c}}&
\colhead{Detections\tablenotemark{d}}&
\colhead{Reference}\\
\colhead{}&
\colhead{(J2000)}&
\colhead{(J2000)}&
\colhead{}&
\colhead{}&
\colhead{Name}&
\colhead{Identifier}&
\colhead{Class}&
\colhead{Observation}&
\colhead{Merit}&
\colhead{}&
\colhead{}
}
\startdata
  J0000.8$+$6600c  & 0.209 & 66.002 & 117.812 & 3.635 & \nodata & \nodata & \nodata & \nodata & \nodata & \nodata & \nodata \\
  J0001.9$-$4158  & 0.483 & $-$41.982 & 334.023 & $-$72.029 & \nodata & \nodata & \nodata & \nodata & \nodata & \nodata & \nodata \\
  J0003.1$+$6227  & 0.798 & 62.459 & 117.388 & 0.108 & \nodata & \nodata & \nodata & \nodata & \nodata &  ROSAT  &  15  \\
  J0004.3$+$2207  & 1.081 & 22.123 & 108.757 & $-$39.448 & \nodata & \nodata & \nodata & \nodata & \nodata & \nodata & \nodata \\
  J0005.1$+$6829  & 1.283 & 68.488 & 118.689 & 5.999 & \nodata & GB6 J0008$+$6837  &  AGN  &  VLA  & 0.804 &  ROSAT  &  15  \\
  J0006.9$+$4652  & 1.746 & 46.882 & 115.082 & $-$15.311 & \nodata & \nodata & \nodata & \nodata & \nodata &  ROSAT  &  15  \\
  J0008.3$+$1452  & 2.084 & 14.882 & 107.655 & $-$46.708 & \nodata & RX J0008.0$+$1450  &  AGN  &  VLA  & 0.700 & \nodata &  3  \\
  J0009.1$+$5031  & 2.289 & 50.520 & 116.089 & $-$11.789 & \nodata & NVSS J000922$+$503028 &  AGN  &  VLA  & 0.941 & \nodata & \nodata \\
  J0016.6$+$1706  & 4.154 & 17.108 & 111.135 & $-$44.964 & \nodata & GB6 J0015$+$1700 &  FSRQ  & \nodata & \nodata & \nodata &  3  \\
  J0017.7$-$0019  & 4.429 & $-$0.326 & 104.735 & $-$62.001 & \nodata & S3 0013$-$00 &  FSRQ  &  VLA  & 0.570 & \nodata &  3  \\
\enddata

\tablecomments{Details of the follow-up observations used to determine the new associations. Where possible, a `figure of merit' was calculated for new AGN associations. Also listed are detections in other waveband that did not lead to new associations. References are given for associations based on other publications. {\bf This table is published in its entirety in the electronic edition of the Astrophysical Journal. A portion is shown here for guidance regarding its form and content.}}

\tablenotetext{a}{Designator for an EGRET source that is positionally associated.}
\tablenotetext{b}{Indicates what (if any) additional data was considered in determining the new association.}
\tablenotetext{c}{A figure-of-merit (FOM) value for sources that were associated using the technique defined in the 1LAC paper \citep{1LAC}; an ellipsis indicates an affiliation, which does not have enough information to calculate a FOM.}
\tablenotetext{d}{Waveband or catalog with association candidates.}

\tablerefs{
(1) \citet{LAT_W28};
(2) \citet{LATW49B};
(3) \citet{1LAC};
(4) \citet{LAT1825HESS};
(5) \citet{HESSsurvey06};
(6) \citet{HESSsurvey08};
(7) \citet{4thIBIScat};
(8) \citet{Castro2010};
(9) \citet{SwiftBATcat};
(10)  \citet{Ghirlanda10};
(11)  \citet{Mahony10};
(12) \citet{Mirabal2009_arXiv};
(13) \citet{Mirabal2010_arXiv};
(14) \citet{Stephen2010};
(15) \citet{Voges94,Voges99,Voges00,Voges2000};
(16) \citet{XMMserendip}
}

\end{deluxetable}
\end{landscape}

\clearpage

\begin{deluxetable}{lrlll}
\tabletypesize{\scriptsize}
\tablecaption{Candidate VHE counterparts
\label{tbl:TeV_candidates}}
\tablewidth{0pt}
\tablehead{
\colhead{1FGL Name}&
\colhead{VHE Source Name}&
\colhead{VHE Association}&
\colhead{Reference}}

\startdata
 J0648.8+1516       & VER J0648+152       & unidentified    & \citet{ATEL2486} \\
 J1503.4$-$5805   & HESS J1503$-$582 & unidentified    & \citet{HESS_J1503-582_2008,HESS_Unid_arxiv} \\
 J1614.7$-$5138c & HESS J1614$-$518 & {\it Pismis 22} & \citet{HESSsurvey06} \\
 J1702.4$-$4147   & HESS J1702$-$402 & PWN of PSR J1702$-$4128 &  \citet{HESSsurvey08} \\
 J1707.9$-$4110c & HESS J1708$-$410 & unidentified    & \citet{HESSsurvey06,HESSsurvey08} \\
 J1839.1$-$0543   & HESS J1841$-$055 & multiple sources & \citet{HESSsurvey08} \\
 J1837.5$-$0659c & HESS J1837$-$069 & unidentified    & \citet{HESSsurvey06} \\
 J1844.3$-$0309   & HESS J1843$-$033 & unidentified    & \citet{Hoppe2008ICRC} \\
 J1848.1$-$0145c & HESS J1848$-$018 & {\it W43}          & \citet{HESS_Unid_arxiv} \\
\enddata

\tablecomments{Candidate VHE counterparts and their associations. Uncertain associations are in italics.}

\end{deluxetable}

\clearpage


\end{document}